\documentclass[prd,preprint,tightenlines,superscriptaddress,floatfix,preprintnumbers,nofootinbib,eqsecnum]{revtex4}

\usepackage{graphicx}
\usepackage{dcolumn}
\usepackage{bm}
\usepackage{soul}
\usepackage{xcolor}
\DeclareUnicodeCharacter{03B3}{$\gamma$}

\def\Pom{{\bf I\!P}}
\def\Reg{{\bf I\!R}}

\def\lsim{\mathrel{\rlap{\lower4pt\hbox{\hskip1pt$\sim$}}
    \raise1pt\hbox{$<$}}}         
\def\gsim{\mathrel{\rlap{\lower4pt\hbox{\hskip1pt$\sim$}}
    \raise1pt\hbox{$>$}}}         

\begin{document}


\title[Photon-photon scattering with FoCal and ALICE 3]{Light-by-light scattering in ultraperipheral collisions of heavy ions 
with future FoCal and ALICE 3 detectors}

\homepage[This paper is dedicated to the memory of Gerard Baur, one of the founders of the UPC physics, who passed away recently.]{}

\author{Paweł Jucha}
\email{Pawel.Jucha@ifj.edu.pl}
 \affiliation{ 
Institute of Nuclear Physics Polish Academy of Sciences,\\
ul. Radzikowskiego 152, PL-31-342 Kraków, Poland
}%
\author{Mariola Kłusek-Gawenda}
\email{Mariolola.Klusek@ifj.edu.pl}
 \affiliation{ 
Institute of Nuclear Physics Polish Academy of Sciences,\\
ul. Radzikowskiego 152, PL-31-342 Kraków, Poland
}

\author{Antoni Szczurek}
\email{Antoni.Szczurek@ifj.edu.pl}
 \affiliation{ 
Institute of Nuclear Physics Polish Academy of Sciences,\\
ul. Radzikowskiego 152, PL-31-342 Kraków, Poland
}%
\affiliation{
College of Mathematics and Natural Sciences, University of Rzeszów,\\
ul. Pigonia 1, PL-35-310 Rzeszów,Poland
}

\date{\today}

\begin{abstract}
We discuss possible future studies of photon-photon
(light-by-light) scattering using a planned FoCal and ALICE 3 detectors. 
We include different mechanisms of $\gamma\gamma\to\gamma\gamma$ scattering such as double-hadronic photon fluctuations,
$t/u$-channel neutral pion exchange or resonance excitations 
($\gamma \gamma \to R$) and deexcitation ($R \to \gamma \gamma$).
The broad range of (pseudo)rapidities and lower cuts on transverse momenta
open a necessity to consider not only dominant box contributions but also
other subleading contributions. 
Here we include low mass resonant $R = \pi^0$, $\eta$, $\eta'$ contributions.
The resonance contributions give intermediate photon transverse momenta.
However, these contributions can be eliminated by imposing windows on
di-photon invariant mass.  
We study and quantify individual box contributions (leptonic, quarkish).
The electron/positron boxes dominate at low $M_{\gamma \gamma}~<$~1~GeV
di-photon invariant masses. 
The PbPb$\to$PbPb$\gamma \gamma$ cross section is calculated
within equivalent photon approximation in the impact parameter space.
Several differential distributions are presented and discussed. We consider four
different kinematic regions. We predict cross section in the (mb-b) range for typical ALICE 3 cuts, a few orders 
of magnitude larger than for the current ATLAS or CMS experiments.
We also consider the two-$\pi^0$ background which can, in principle, 
be eliminated at the new kinematical range
for the ALICE 3 measurements by imposing dedicated cuts on di-photon transverse momentum and$\slash$or so-called vector asymmetry.
\end{abstract}

\maketitle


\section{\label{sec:level1}  Introduction}


The photon-photon scattering, often called light-by-light scattering,
is an interesting quantum effect. Till recently, it was not studied
experimentally. It was a dream for the laser community. The works in this
direction by the laser community are going on \cite{laser}. The possibility of light-by-light 
studies in ultraperipheral heavy ion collisions 
was proposed in \cite{Enterria,PhysRevC.93.044907}. 
Inspired by the theoretical analyses it was then studied experimentally 
by the ATLAS \cite{ATLAS1} and CMS \cite{CMS1} collaborations.
Statistics improved data were presented in \cite{ATLAS2,ATLAS3}.
The experimental data can be almost explained taking into account
only QED box contributions.
The ATLAS and CMS measurements can register only transverse momenta 
of photons larger than about $2-2.5$ GeV, i.e. automatically large diphoton masses.
As matter of course, this means small statistics of several tens of events.
It was discussed in \cite{KSS} what the ALICE and LHCb collaborations could do 
for smaller di-photon invariant masses. According to our knowledge, 
the experimental analysis of the ALICE collaboration is in progress.
The previous studies of nuclear reactions considered almost exclusively so-called box contributions.
Other underlying mechanisms were discussed rather only  for $\gamma\gamma\to\gamma\gamma$ 
scattering in \cite{PhysRevC.93.044907,KSS,Nasza3_2gluo,LP}.
The authors of \cite{Coelho:2020syp} considered also diffractive mechanisms 
of production of two photons associated however with extra hadronic emissions.
The $\gamma\gamma \to \gamma\gamma$ is also interesting in the context of searching 
for effect beyond Standard Model \cite{Baldenegro:2018hng}.

In this analysis, we explore what future FoCal \cite{FoCAL} and ALICE 3 \cite{ALICE3} detectors
could do in this respect. A forward electromagnetic calorimeter is planned as an upgrade to the ALICE experiment for data-taking in 2027-2029 at the LHC. The FoCal will cover pseudorapidities range of $3.4<\eta<5.8$. Runs 5 and 6 will allow to measure more than five times the present Pb-Pb luminosity. This increase of luminosity, in combination with improved detector capabilities, will enable the success of the physical program planned in ALICE 3. A significant feature of FoCal and ALICE 3 programs is the ability to measure photons in relatively low (starting from a few MeV) transverse momenta. 

In the present paper we will consider not only box contributions
but also contributions of the other mechanisms (double photon hadronic
fluctuations, $\pi^0$ t$\slash$u-channel exchanges, two-gluon exchange, etc.).
We will explore whether the other mechanisms that are less under
theoretical control (non-perturbative pQCD domain), can be observed 
experimentally with the future apparatus. We will try to find conditions how to relatively 
enhance them compared to the box contributions to be observed 
in heavy ion UPC. 

\section{\label{sec:level2}  Sketch of the formalism}

\subsection{Elementary cross section and general remarks}

The Weizsäcker-Williams formula (see subsection \ref{subsec:nuclear_cs}) is based on the knowledge of the elementary $\gamma\gamma \to \gamma\gamma$ cross section. The angular distribution of $\gamma\gamma\to\gamma\gamma$ depends on $z=\cos(\theta)$ where $z$ is in range $(-1,1)$, and $\theta$ is the scattering angle, and the mass of the particles produced in the process $M_{\gamma\gamma}$. Determination of the elementary cross section requires a calculation of photon-photon scattering amplitudes derived from Feynman diagrams of fermion loops. As shown in \cite{Nasza3_2gluo}, higher-order processes such as VDM-Regge and 2-Gluon exchange can be important at energies above 30 GeV, while in the low-energy regime, these processes should not play a significant role.

The distribution of the elementary cross section for low energies is known and has been shown at least in \cite{PhysRevC.93.044907}, but so far, no one has used this for light-by-light scattering calculation below 5 GeV due to experimental limitations. Moreover, this is the first paper where the influence of the different types of particles generated in the loop on total cross section is shown explicitly.

In the present work, the minimum mass of invariant produced photons is $10$~MeV, which, according to the formula $p_{t,min} = M_{min}/2$, means that the minimum value of the transverse momentum can be 5 MeV. The elementary cross section here is calculated for unpolarised photons. To do this, all $16$ photon helicity combinations of the cross section$\slash$amplitude must be added up. In fact, due to symmetries, it is sufficient to count only five combinations and then add them up with the corresponding weights:
 \begin{eqnarray}
     \sum_{{\lambda_1,\lambda_2,\lambda_3,\lambda_4}} \left|\mathcal{A}^{\gamma \gamma \rightarrow \gamma \gamma}_{\lambda_1\lambda_2 \to \lambda_3 \lambda_4}\right|^2 &=& 2\left|\mathcal{A}_{++++}^{fermions}\right|^2 + 2\left|\mathcal{A}_{+--+}^{fermions}\right|^2 \nonumber \\ &+&2\left|\mathcal{A}_{+-+-}^{fermions}\right|^2 + 2\left|\mathcal{A}_{++--}^{fermions}\right|^2 
     +8\left|\mathcal{A}_{+-++}^{fermions}\right|^2 \;.
     \label{helic}
 \end{eqnarray}
Elementary cross section calculations for the box contribution were carried out using FormCalc and LoopTools libraries based on Mathematica software.


\subsection{Double-photon hadronic fluctuations}

This component was calculated for the first time in \cite{PhysRevC.93.044907, Nasza3_2gluo}
assuming vector dominance model. In this approach, the amplitude 
for the process is given as:
\begin{eqnarray}
{\cal M} &=& \Sigma_{i,j} C_i^2 C_j^2 \left( C_{\Pom} \left( \frac{s}{s_0} \right)^{\alpha_{\Pom}(t)-1} F(t)
                                + C_{\Reg} \left( \frac{s}{s_0} \right)^{\alpha_{\Reg}(t)-1}F(t) \right)
\; ,
\nonumber \\
         &+& \Sigma_{i,j} C_i^2 C_j^2 \left( C_{\Pom} \left( \frac{s}{s_0} \right)^{\alpha_{\Pom}(u)-1} F(u)
                                + C_{\Reg} \left( \frac{s}{s_0} \right)^{\alpha_{\Reg}(u)-1} F(u) \right)
\; .
\label{SS_amplitude}
\end{eqnarray}
In the simplest version of the model $i, j = \rho^0, \omega, \phi$ 
(only light vector mesons are included).
The couplings $C_i, C_j$ describe the $\gamma \to V_{i/j}$ transitions
that are calculated based on vector meson dilepton width.
$C_{\Pom}$ and $C_{\Reg}$ are extracted from the Regge 
factorization hypothesis (see e.g. \cite{SNS,SS}).

It was shown in \cite{PhysRevC.93.044907} that the component is concentrated mainly 
at small photon transverse momenta which at not too small subsystem
energies corresponds to $z \approx \pm$ 1.
The Regge trajectories are usually written in a linear form:
\begin{eqnarray}
\alpha_{\Pom}(t/u) = \alpha_{\Pom}(0) + \alpha_{\Pom}'t/u \; , \nonumber \\
\alpha_{\Reg}(t/u) = \alpha_{\Reg}(0) + \alpha_{\Reg}'t/u \; .
\label{linear_trajectories}
\end{eqnarray}
These linear forms are valid at not too large $|t|$ or $|u|$.
At large $|t|$ or $|u|$ the energy dependent factors are artificially small.
Therefore here where we explore it more, we propose to smoothly
switch off the $t/u$ dependent terms in (\ref{linear_trajectories})
at $t \sim$ -0.5 GeV$^2$ and $u \sim$ -0.5 GeV$^2$. 
The actual place where it should be done is not known precisely.
Another option would be to use $\sqrt{t/u}$ trajectories 
\cite{Brisudova2, Brisudova1}. 

We also wish to analyze whether more heavy vector mesons such as 
$J/\psi$ can give a sizeable contribution.

For example, for the double $J/\psi$ fluctuations 
(both photons fluctuate into virtual $J/\psi$ mesons) 
we take the following Ansatz for the helicity
conserving amplitude:
\begin{eqnarray}
{\cal M}_{VDM}^{J/\psi J/\psi} &=& g_{J/\psi}^2 C_{\Pom}^{J/\psi} 
\left( \frac{s}{s_0} \right)^{\alpha_{\Pom}^{J/\psi J/\psi}(t) - 1}
  F_{J/\psi J/\psi \Pom}^{H}(t) F_{J/\psi J/\psi \Pom}^{H}(t)  
\nonumber\\
                               &+& g_{J/\psi}^2 C_{\Pom}^{J/\psi}
\left( \frac{s}{s_0} \right)^{\alpha_{\Pom}^{J/\psi J/\psi}(u) - 1}
  F_{J/\psi J/\psi \Pom}^{H}(u) F_{J/\psi J/\psi \Pom}^{H}(u)
\; .
\label{HH_amplitude}
\end{eqnarray}
In this case (double $J/\psi$ fluctuations) only pomeron can be exchanged 
(no subleading reggeons are possible due to the $c \bar c$ structure of 
$J/\psi$ mesons).
In this case, for simplicity, we take the simplified trajectories as
\begin{equation}
\alpha_{\Pom}^{J/\psi J/\psi}(t) = \alpha_{\Pom}^{J/\psi J/\psi}(u) =
\alpha_{\Pom}^{J/\psi J/\psi}(0) \; .
\label{trajectory_forJpsiJpsi}
\end{equation}
Here the $t/u$ dependencies of the trajectories are totally ignored.
In numerical calculations we take $\alpha_{\Pom}^{J/\psi J/\psi}(0) =
1.3 - 1.4$ (typical hard pomeron).
Since the $J/\psi$ mesons are far off-mass-shell and more compact than light
vector mesons also the form factors must be modified.
Here we take them in the form:
\begin{eqnarray}
F_{J/\psi J/\psi \Pom}^H(t) = \exp\left( \frac{t-m_{J/\psi}^2}{\Lambda_{J/\psi}^2}
\right) \; , \\
F_{J/\psi J/\Psi \Pom}^H(u) = \exp\left( \frac{u-m_{J/\psi}^2}{\Lambda_{J/\psi}^2}
\right) \; .
\label{hard_formfactors}
\end{eqnarray}
Please note that the form factors are normalized to 1 on the meson ($J/\psi$)
mass shell. One could also use monopole-like form factors.
These form factors drastically reduce the $J/\psi J/\psi$
component of the amplitude in comparison to light vector meson components.
However, due to compactness of $J/\psi$ we expect $\Lambda_{J/\psi}$ to 
be large. 
In the calculations presented here, we take $\Lambda_{J/\psi}=2$~GeV
for illustration. The actual number is not well known.
Also the normalization parameter $C_{\Pom}^{J/\psi}$ is not well known.
It is expected to be smaller than for light vector mesons.

In a similar fashion, one could include one $J/\psi$ fluctuation and one
light vector meson fluctuation. However, there the choice of trajectories
is unclear. We will leave these components for future detailed studies.

Finally, let us discuss the helicity structure of the double photon
hadronic fluctuation amplitude.
We write:
\begin{eqnarray}
{\cal M}_{\lambda_1 \lambda_2 \to \lambda_3 \lambda_4}^{(t)} &=&
A(t) \; \delta_{\lambda_1 \lambda_3} \delta_{\lambda_2 \lambda_4} \; , \\
{\cal M}_{\lambda_1 \lambda_2 \to \lambda_3 \lambda_4}^{(u)} &=&
A(u) \; \delta_{\lambda_1 \lambda_4} \delta_{\lambda_2 \lambda_3} \; .
\label{amplitudes}
\end{eqnarray}
$A(t)$ and $A(u)$ are given explicitly in (\ref{SS_amplitude}).
Then the total double VDM amplitude, including $t$ and $u$ processes, reads:
\begin{equation}
{\cal M}^{VDM}_{\lambda_1 \lambda_2 \to \lambda_3 \lambda_4} =
\frac{1}{\sqrt{2}} \left(
{\cal M}^{VDM,(t)}_{\lambda_1 \lambda_2 \to \lambda_3 \lambda_4} +
{\cal M}^{VDM,(u)}_{\lambda_1 \lambda_2 \to \lambda_3 \lambda_4}
\right)
\; .
\end{equation}
Having the double VDM helicity amplitudes, we can add different mechanisms
coherently:
\begin{equation}
{\cal M}_{\lambda_1 \lambda_2 \to \lambda_3 \lambda_4} =
{\cal M}^{boxes}_{\lambda_1 \lambda_2 \to \lambda_3 \lambda_4} +
{\cal M}^{VDM}_{\lambda_1 \lambda_2 \to \lambda_3 \lambda_4} +
{\cal M}^{\pi^0}_{\lambda_1 \lambda_2 \to \lambda_3 \lambda_4} + ... \; .
\label{summing_amplitudes}
\end{equation}
In the following, we shall discuss the sum of the larger two components 
(boxes and VDM) and quantify their interference effects.


\subsection{Nuclear cross section}
\label{subsec:nuclear_cs}

In the present paper, the nuclear cross section is calculated using equivalent photon approximation (EPA) in the b-space. In this approach, the di-photon cross section can be written as (see \cite{EPA_eq}):

\begin{eqnarray}
    \frac{d\sigma(PbPb \to PbPb \gamma \gamma)}{dy_{\gamma_1}dy_{\gamma_2}dp_{t,\gamma}} &=& 
    \int \frac{d\sigma_{\gamma\gamma\to\gamma\gamma}(W_{\gamma\gamma})}{dz}N(\omega_1,b_1)N(\omega_2,b_2)S^2_{abs}(b) \nonumber \\  
    &\times& d^2b d\bar{b_x} d\bar{b_y} \frac{W_{\gamma\gamma}}{2} \frac{dW_{\gamma\gamma}dY_{\gamma\gamma}}{dy_{\gamma_1}dy_{\gamma_2}dp_{t,\gamma}} dz \;,
    \label{eq:tot_xsec}
\end{eqnarray}
where $\bar{b}_x = \left( b_{1x} + b_{2x}\right)/2$ and $\bar{b}_y = \left( b_{1y} + b_{2y}\right)/2$. The relation between $\vec{b}_1$, $\vec{b}_2$ and impact parameter: $b = |\vec{b}| = \sqrt{|\vec{b}_1|^2 + |\vec{b}_2|^2 - 2|\vec{b}_1||\vec{b}_2|\cos\phi}$. Absorption factor $S^2_{abs}(b)$ is calculated as:

\begin{eqnarray}
    S^2_{abs}(b) = \Theta(b-b_{max})  
\end{eqnarray}
%
or
\begin{eqnarray}
    S^2_{abs}(b) = exp\left( -\sigma_{NN} T_{AA}(b) \right) \;,
    \label{eq:s2b}
\end{eqnarray}
%
where $\sigma_{NN}$ is the nucleon-nucleon interaction cross section, and $T_{AA}(b)$ is related to the so-called nuclear thickness, $T_A(b)$,
\begin{equation}
    T_{AA}\left(|\vec{b}| \right) = \int d^2\rho T_A \left( \vec{\rho} -\vec{b} \right) T_A\left(\rho\right) ,
\end{equation}
%
and the nuclear thickness is obtained by integrating the nuclear density 
\begin{equation}
    T_A \left( \vec{\rho} \right) = \int \rho_A\left( \vec{r} \right) dz , \hspace{0.5cm} \vec{r} = \left( \vec{\rho},z \right) \;,
\end{equation}
%
where $\rho_A$ is the nuclear charge distribution.
The nuclear photon fluxes $N(\omega_1,b_1)$ and $N(\omega_2,b_2)$ are calculated with realistic charge distribution, as described in \cite{KSpi}.

So far, in our previous works, we presented UPC results only with a sharp cut on the impact parameter, which reflects the distance between two nuclei with a value equal to exactly two radii of the nuclei, i.e. $b>14$~fm for Pb+Pb collisions. Due to the no homogeneous nuclear charge distribution, it seems to be more reasonable to use the absorption factor given by Eq.~(\ref{eq:s2b}).

\subsection{Background contribution}

It was discussed in \cite{KSS} that the $\gamma \gamma \to \pi^0(\to2\gamma)\pi^0(\to2\gamma)$ reaction constitutes a difficult background for the $\gamma\gamma \to \gamma\gamma$ measurements at intermediate $M_{\gamma\gamma}$. How to calculate the cross section for $\gamma\gamma \to \pi^0\pi^0$ reaction was discussed in \cite{Klusek-Gawenda:2013rtu} and will be not repeated here. The calculation of the background proceeds in three steps. First the cross section for $\gamma\gamma \to \pi^0\pi^0$ is calculated (for details see \cite{Klusek-Gawenda:2013rtu}). Next the cross section for $AA \to AA\pi^0\pi^0$ is computed in the equivalent photon approximation in an analogous way as described in the previous subsection. Finally the simulation of both $\pi^0$ decays is performed and combined distributions of one photon from the first $\pi^0$ and one photon from the second $\pi^0$ are constructed. 


\section{\label{sec:level3}  Results - Elementary cross section}

\begin{figure}[!h]
	(a)\includegraphics[scale=0.32]{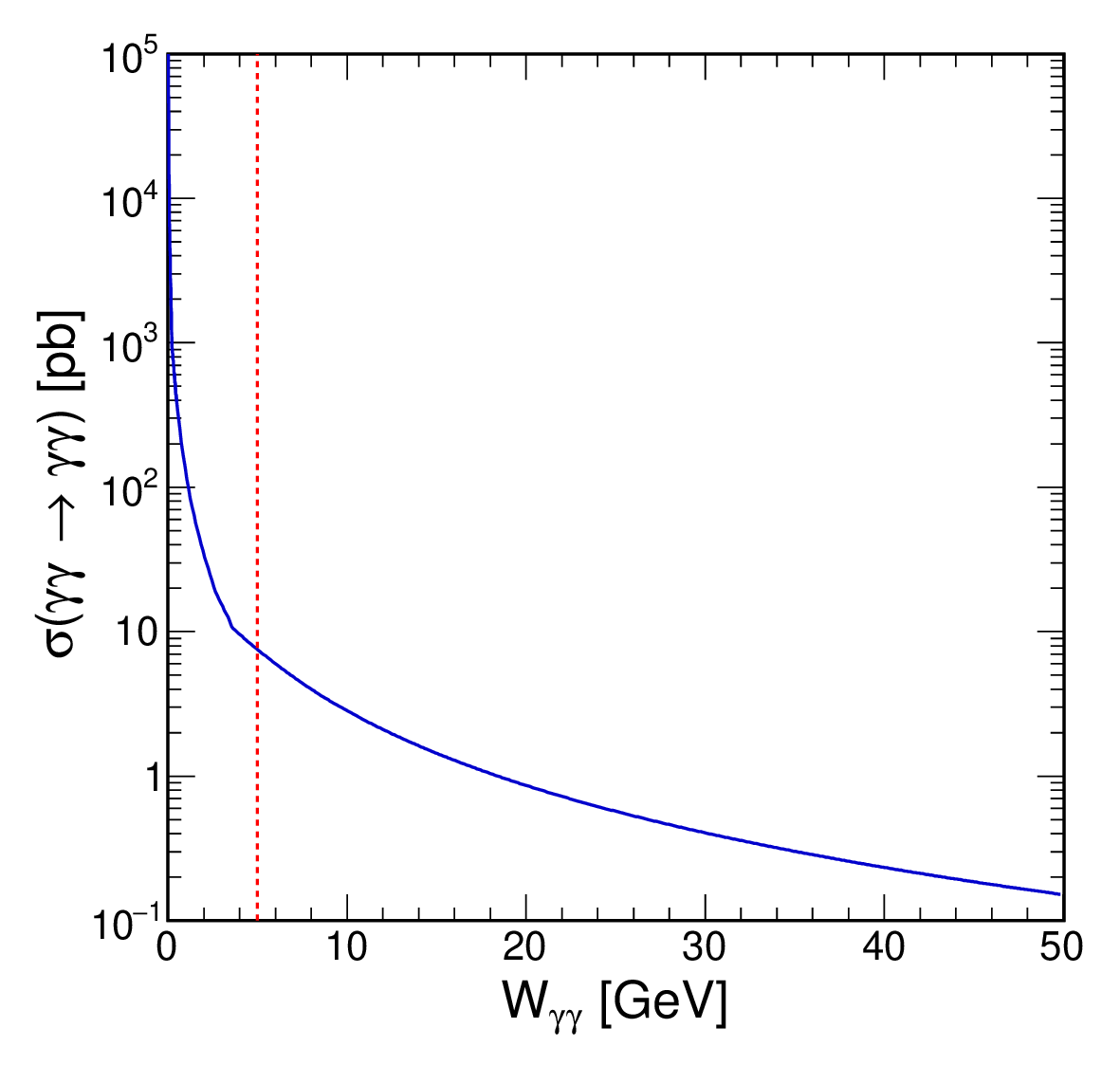}
	(b)\includegraphics[scale=0.32]{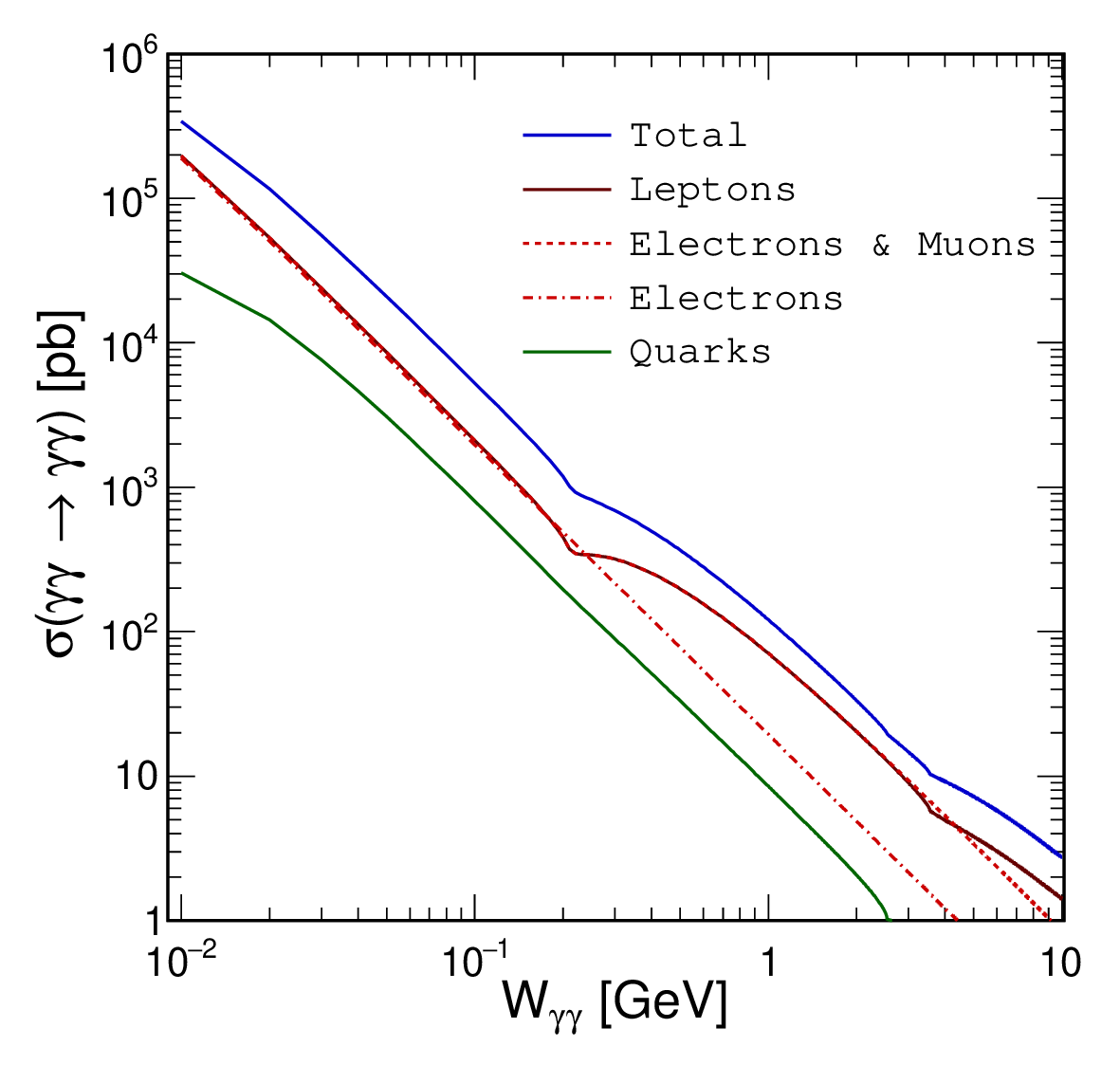}
	\caption{\label{fig:elementary_boxes} 
		Elementary cross section (in pb) as a function of energy. Results for fermionic boxes are shown over a wide range of energies (a) - linear scale and focusing on the MeV range (b) - where logarithmic scale is used. The total cross section (blue solid line) is split somewhat artificially into quarks (green solid line), electrons (red dashed line), electrons and muons (red dotted line), and leptons (red solid line) contributions. }
\end{figure}

Treating photons as massless particles, one can expect the absence of a minimum energy value, which is determined by the kinematical threshold. In Fig.~\ref{fig:elementary_boxes} we show the dependence of the elementary cross section on energy in the $\gamma \gamma$ system.
Fig.~\ref{fig:elementary_boxes}(a) demonstrates that putting a cut on the energy of the two-photon system (the red dashed vertical line indicates the value of $W_{\gamma\gamma}=5$~GeV), we automatically get rid of a significant signal contribution. The pointed limitation is due to the existing restriction of the detectors measuring light-by-light scattering in the ATLAS and CMS experiments. 
Focusing on the details of box contribution,
Fig.~\ref{fig:elementary_boxes}(b), we show individual 
contributions of different boxes (electron, muon, quarks).
One can see that at low diphoton invariant masses, the electronic loops dominate.
The quarkish loops become sizeable only at $W_{\gamma \gamma} >$ 1 GeV. At an energy of $W_{\gamma\gamma}>2m_\tau$ one can observe a slight enhancement of the fermionic contribution, which illustrates the presence of $\tau$ leptons in the loop.

\begin{figure}[!h]
	(a)\includegraphics[scale=0.25]{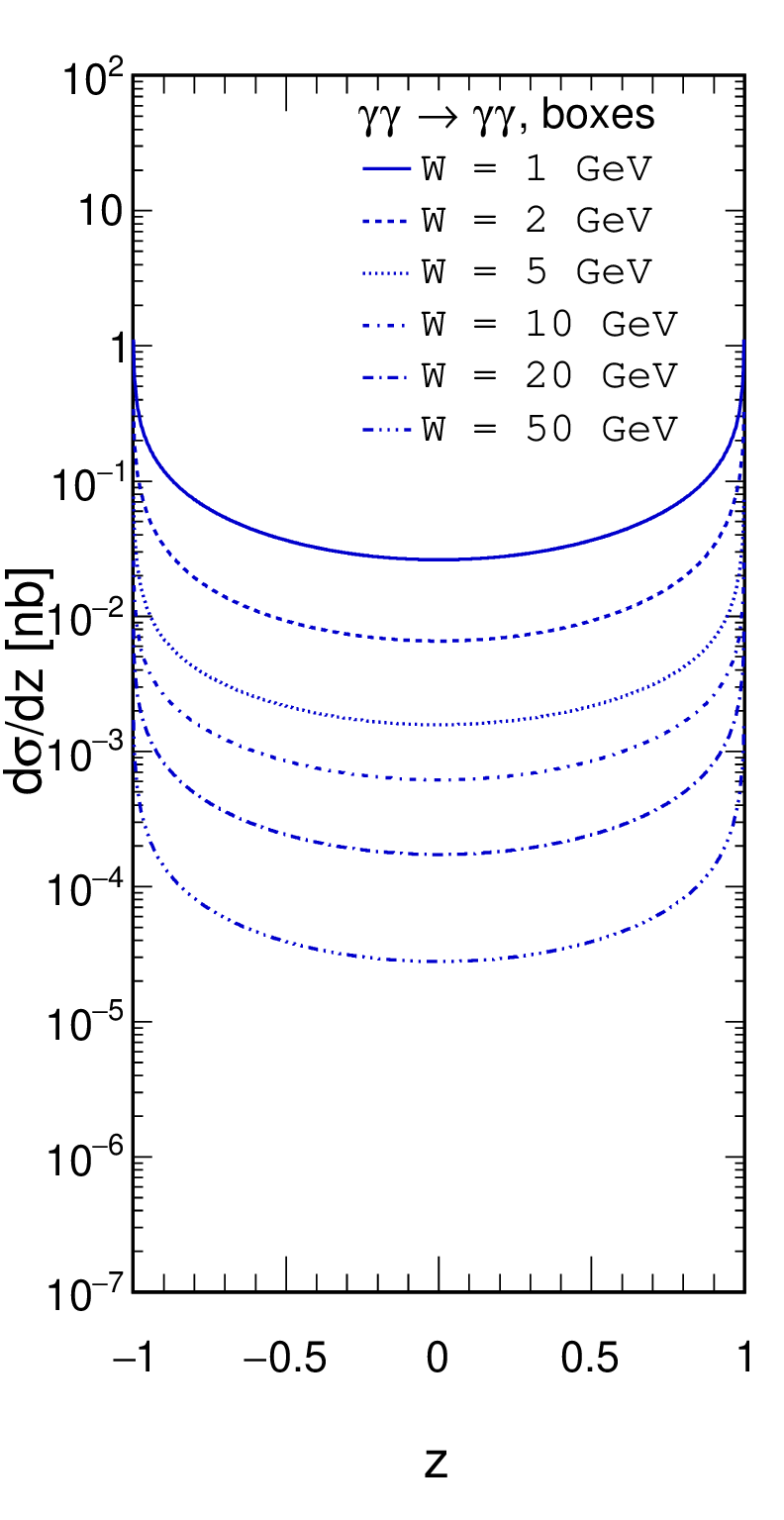}
	(b)\includegraphics[scale=0.25]{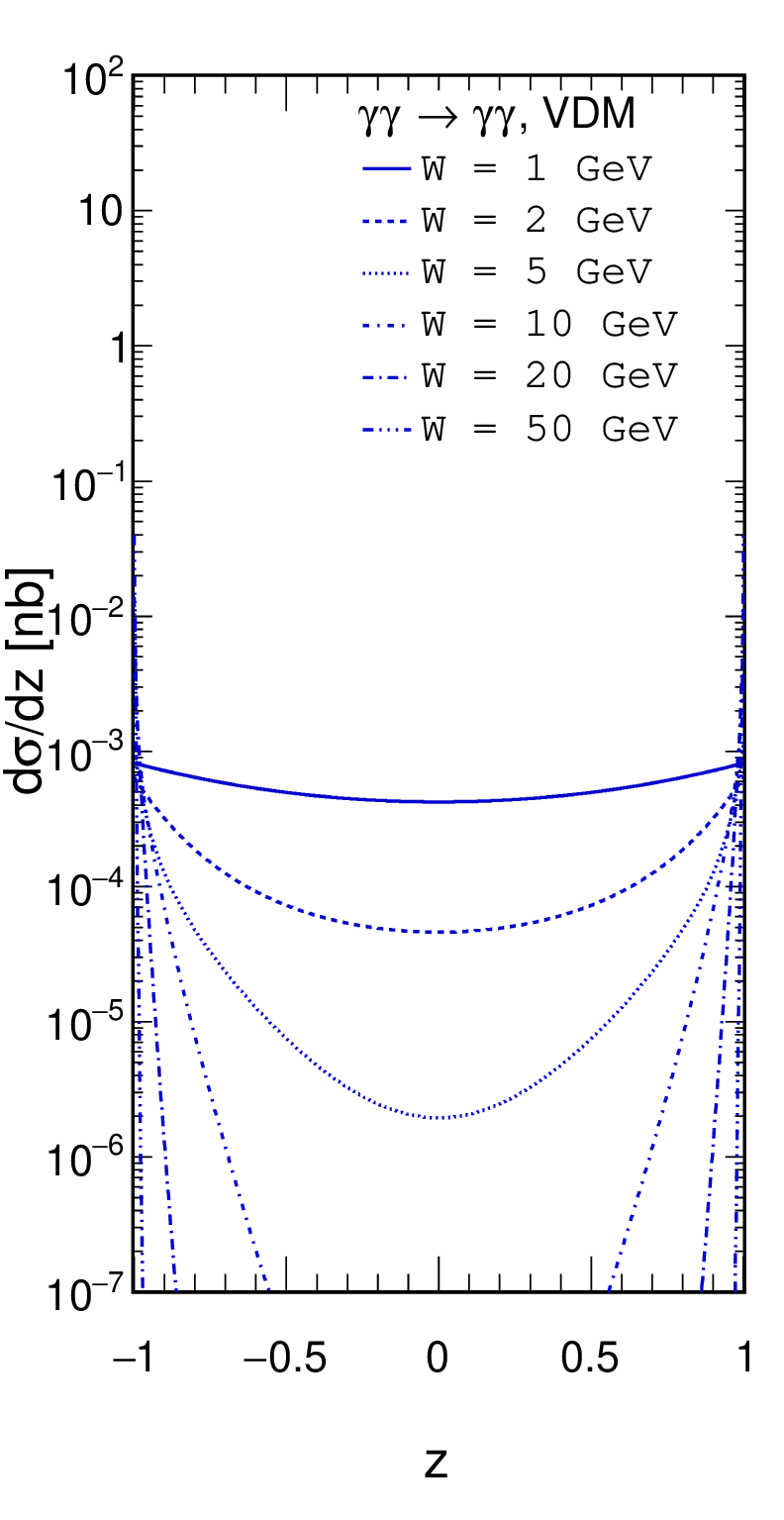}
	(c)\includegraphics[scale=0.25]{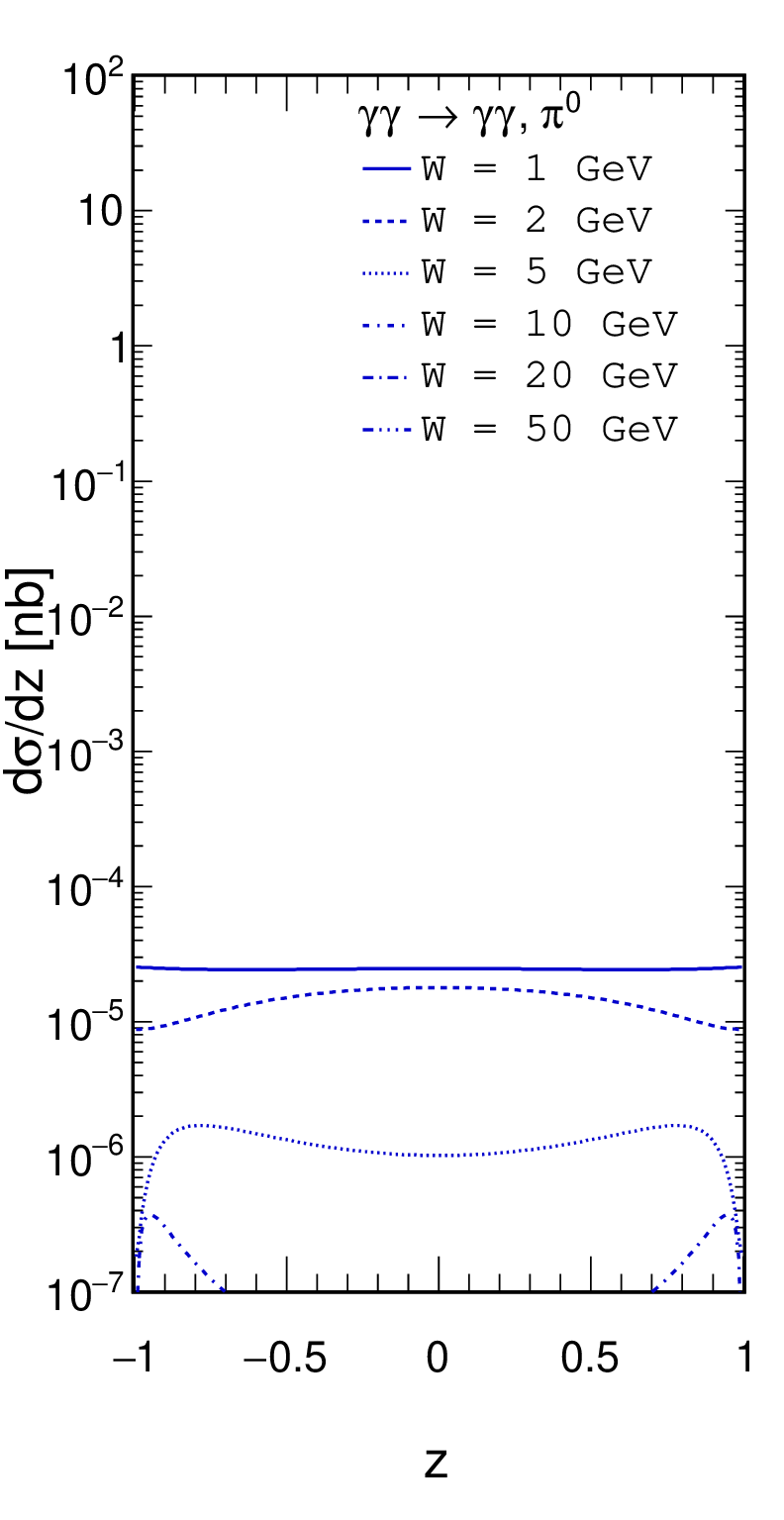}
	\caption{$cos(\theta)$ distributions for (a) boxes, (b) double hadronic fluctuations
	and (c) $\pi^0$ exchange for different
		energies $W= 1, 2, 5, 10, 20, 50$~GeV.
	}
	\label{fig:gamgam_gamgam_subleading}
\end{figure}

In Fig.~\ref{fig:gamgam_gamgam_subleading} we show $d \sigma / dz$ for $\gamma\gamma \to \gamma\gamma$
for (a) boxes, (b) double hadronic fluctuation calculated within the VDM-Regge approach
and (c) the $\pi^0$-exchange calculated as in Ref.~\cite{LS}. Results are presented for five fixed values of energy in the range of $(1-50)$~GeV.
At larger energies the VDM contribution peaks at $z = \pm 1$.
On the other hand, the $\pi^0$ exchange contribution has minima 
at $z = \pm 1$ which is due to the structure of corresponding vertices. The latter contribution is relatively small. In general, the box contributions dominate, especially for low photon-photon scattering energies. At larger scattering energies 
($W_{\gamma\gamma} >2$  GeV) the VDM contribution competes with the box contributions
only at $z \sim \pm 1$.
Can one expect in this context sizeable interference effects of both mechanisms?

\begin{figure}[!h]
(a)\includegraphics[scale=0.32]{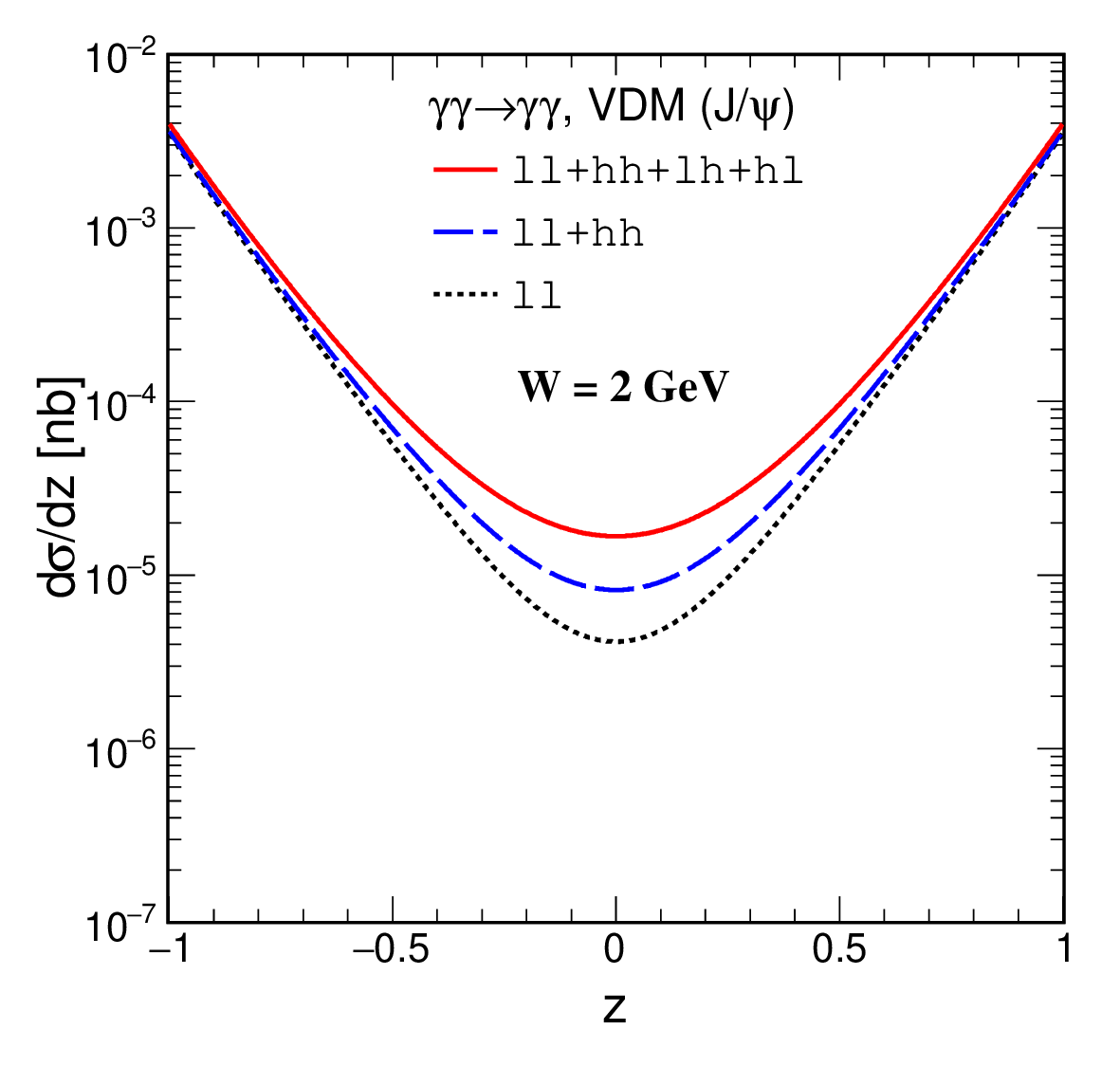}
(b)\includegraphics[scale=0.32]{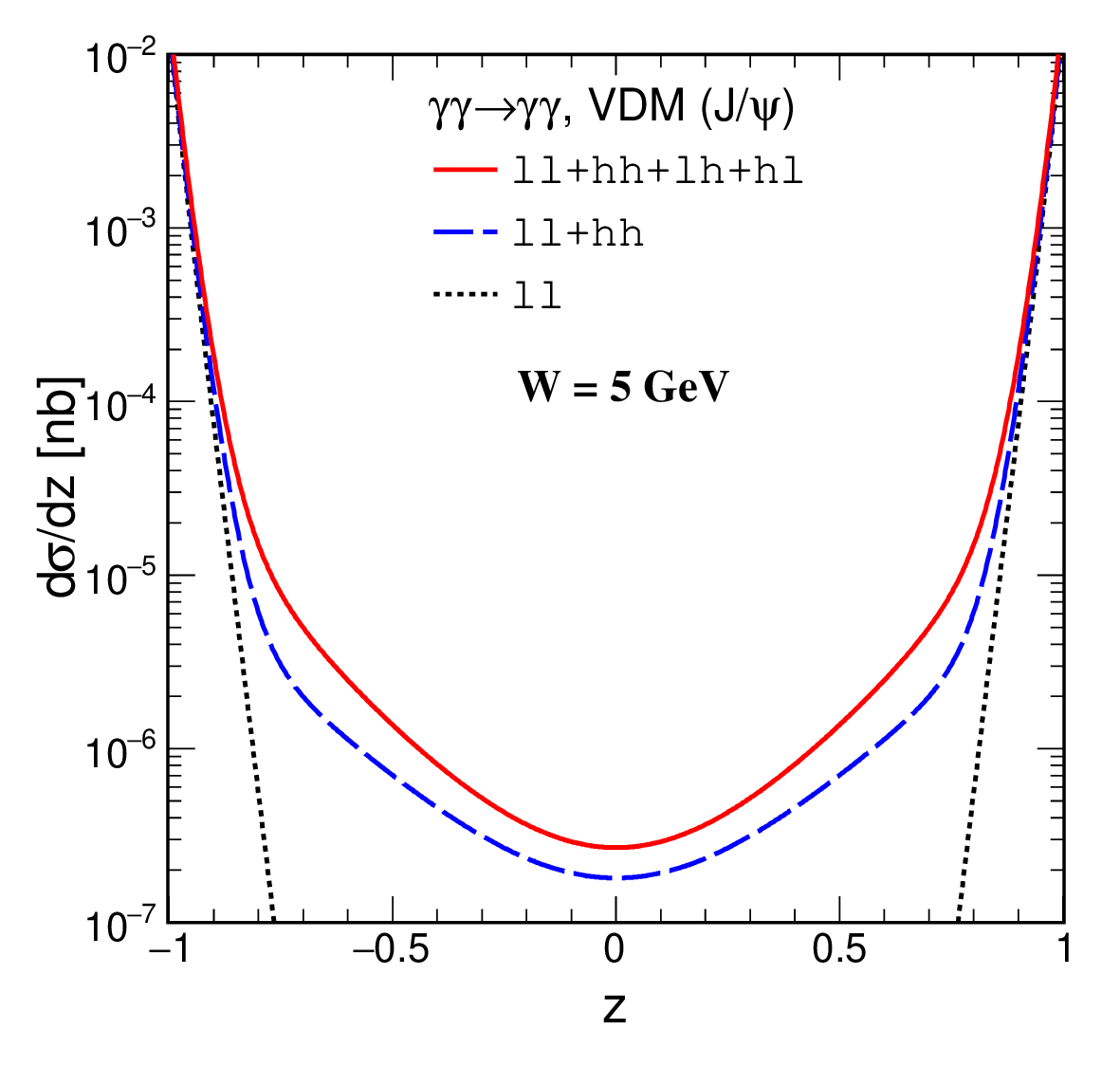}
\caption{Modification of $d\sigma/dz$ due to including
fluctuations with virtual $J/\psi$ mesons: (a) $W = 2$~GeV, (b) $W = 5$~GeV.
The top solid line includes all components (light (\textit{l}) and heavy (\textit{h}) vector mesons), the dotted line only light vector mesons.
}
\label{fig:dsig_dz_vdm_hard} 
\end{figure}

Now we wish to concentrate briefly on second biggest in 
Fig.\ref{fig:gamgam_gamgam_subleading} contribution, 
double photon fluctuations. We include both light vector mesons $\rho^0$,
$\omega$, $\phi$) as well as $J/\psi$ (one or two) as decribed in the
theoretical section.
Our results, for two collision energies ($W = 2, 5$~GeV), are shown in
Fig.\ref{fig:dsig_dz_vdm_hard}. The dotted line includes only light
vector meson fluctuations, the dashed line in addition double 
$J/\psi$ fluctuations and the solid line all combinations of photon 
fluctuations.
Inclusion of  $J/\psi$ meson fluctuations leads to an enhancement of 
the cross section at -0.5 $< z <$ 0.5. The enhancement is more spectacular 
for larger collsion energy. The corresponding cross section there 
is, however, much smaller than the box contribution 
(see Fig.\ref{fig:gamgam_gamgam_subleading}).

\begin{figure}[!h]
	\includegraphics[scale=0.32]{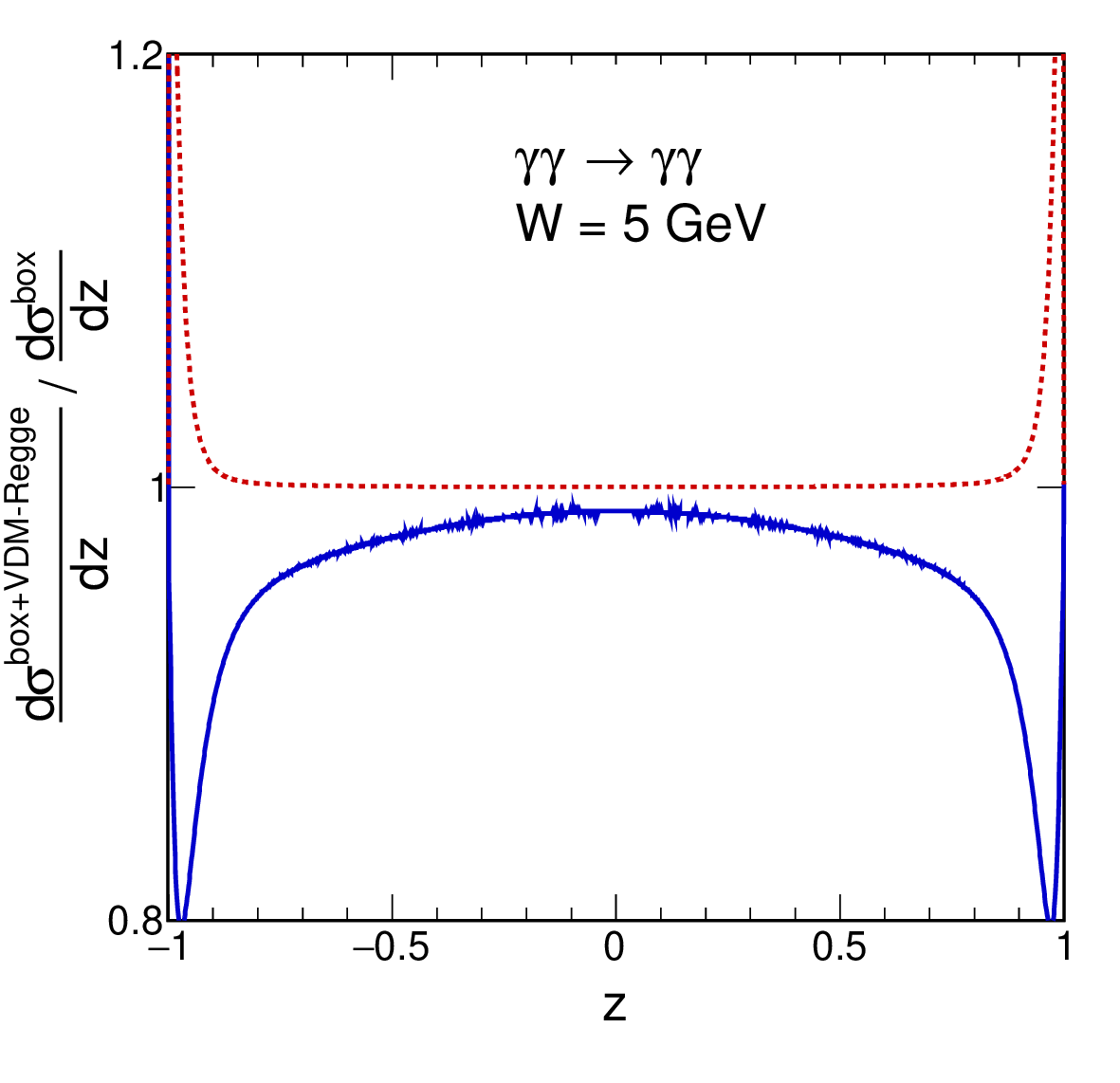}
	\caption{The ratio of the coherent (blue) and incoherent (red) sum 
		of the box and VDM-Regge contributions divided by the cross section for the box contribution alone for $W = 5$~GeV.}
	\label{fig:ratio_z}
\end{figure}

Now we wish to quantify how much is the box result changed when adding
the VDM-Regge contribution. In Fig.~\ref{fig:ratio_z} we show the ratio
of cross sections when including the VDM-Regge contribution to that for
box contribution only. In this calculation, only the helicity
contributions that are active in the VDM component (6 combinations)
are included.
The red line represents the incoherent sum,
while the blue line also includes interference effects.
In this calculation, the so-called ``sqrt'' trajectories \cite{Brisudova1, Brisudova2} were used.
We observe a negative interference effect. Adding the remaining
contributions would lead to additional deviations.

\section{\label{sec:level4}  Results - nuclear cross section}

Now we go to nuclear UPC and will show our results for four experimental kinematic conditions, each in a separate subsection.

\subsubsection{ATLAS and CMS kinematics}

\begin{figure}[!h]
	\begin{minipage}{0.48\columnwidth}
		(a)\includegraphics[scale=0.3]{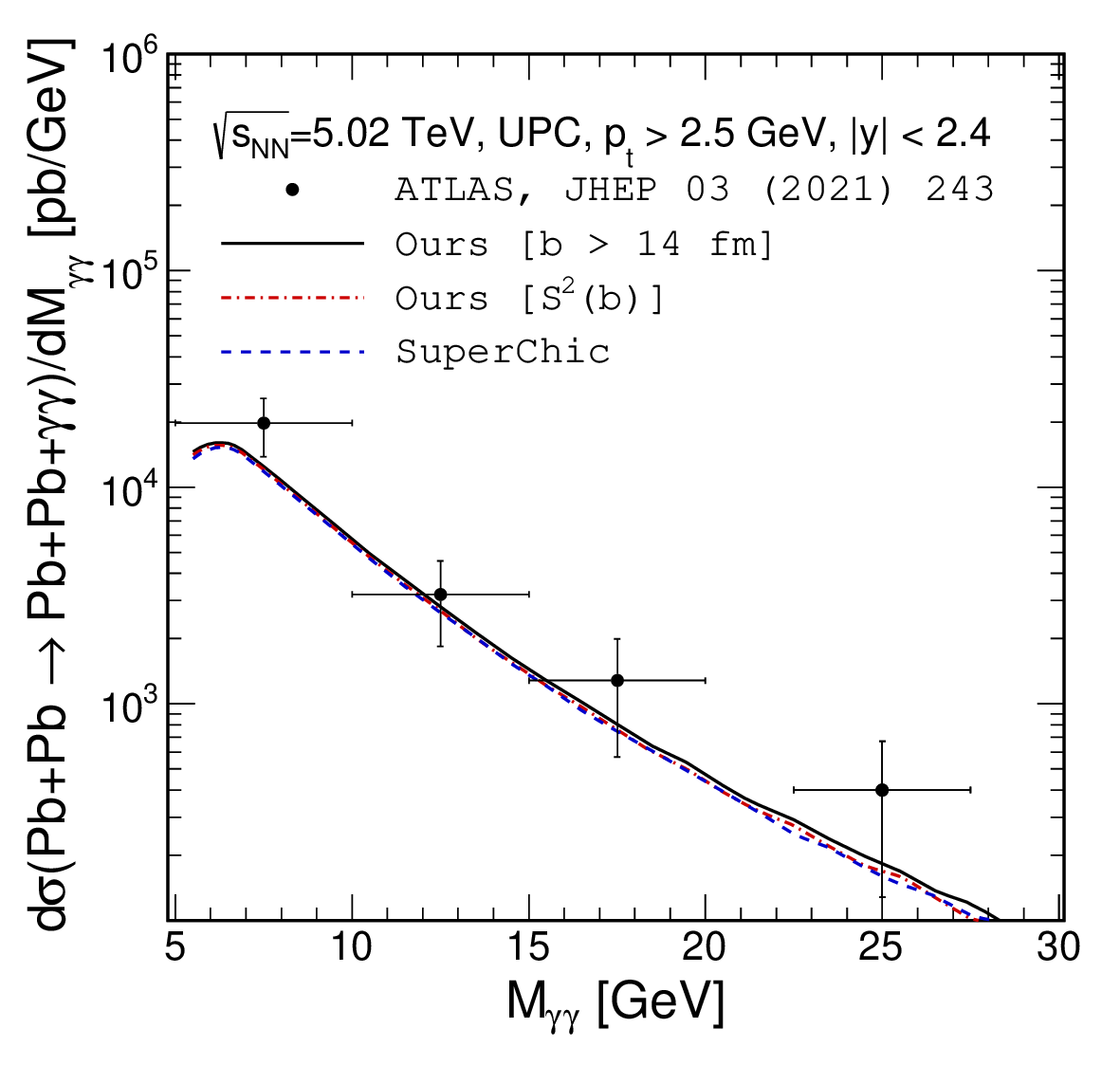}
	\end{minipage}
	\begin{minipage}{0.48\columnwidth}
		(b)\includegraphics[scale=0.3]{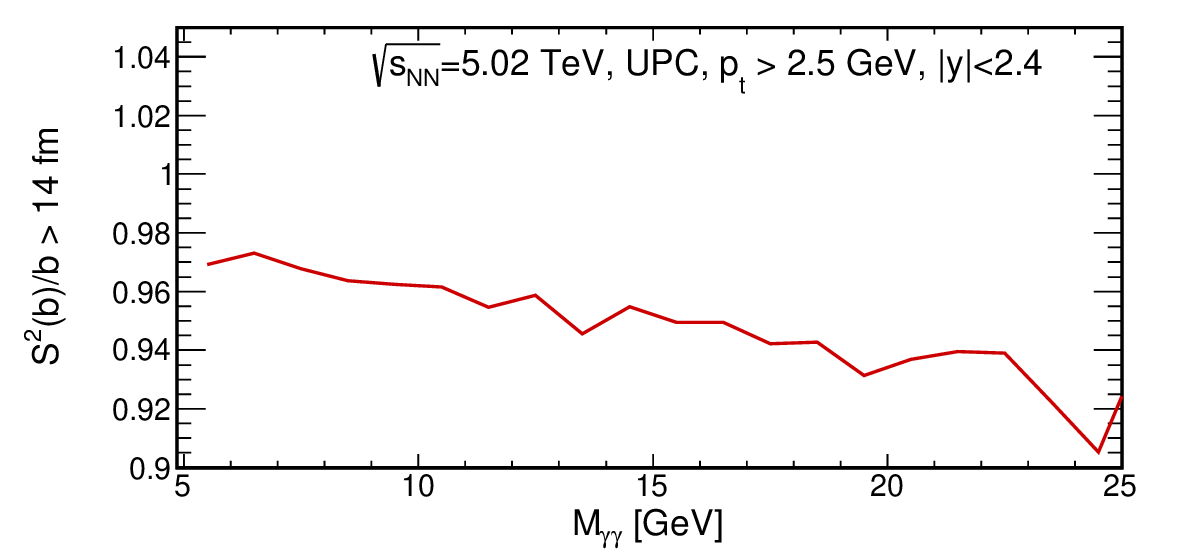}\\
		(c)\includegraphics[scale=0.3]{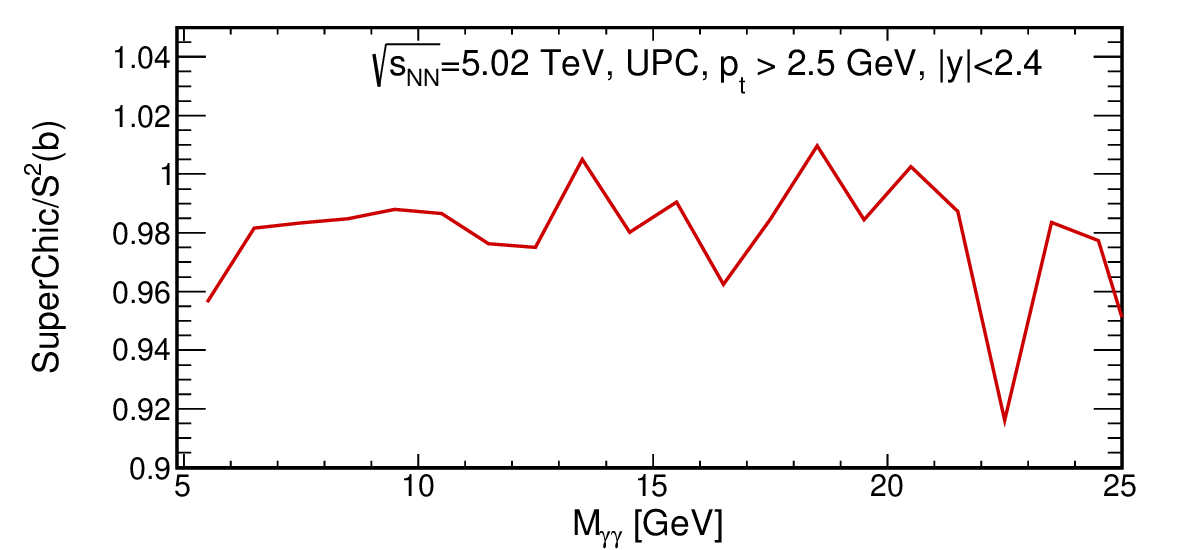}
	\end{minipage}
	\caption{\label{fig:2} Differential cross section as a function of two-photon invariant mass at $\sqrt{s_{NN}}=5.02$~TeV. (a) The ATLAS experimental data are collected with theoretical results including a sharp cut on impact parameter ($b>14$~fm - solid black line) and smooth nuclear absorption factor $S^2(b)$ (dash-dotted red line). For completeness, results that are obtained with the help of Eq.~(\ref{eq:tot_xsec}) are compared with results from SuperChic \cite{HarlandLang2020veo}. The right panel shows two ratios: (b) results from our approach comparing sharp and smooth cut-off on impact parameter and (c) SuperChic outcome to our results, using a smooth representation of the gap survival factor.   }
\end{figure}

We start by confronting our calculations with the current
ATLAS data \cite{ATLAS2}. 
%
%
Fig.~\ref{fig:2}(a) 
shows di-photon invariant mass distribution \cite{ATLAS2}. This result slightly depends on the treatment of absorption corrections
in the $b$-space. The results of the two different approximations (as described in the figure caption) almost coincide. For comparison, we show results obtained with SuperChic generator \cite{HarlandLang2020veo}.
We get a reasonable agreement taking into account relatively large error bars of experimental data  (small statistics). Right panel of Fig.~\ref{fig:2} presents the ratio of nuclear results from our approach (Eq.~(\ref{eq:tot_xsec})) comparing smooth cut-off (Eq.~(\ref{eq:s2b})) and sharp cut on impact parameter ($b>14$~fm) corresponds to results from SuperChic generator \cite{HarlandLang2020veo}. The difference between results including smooth and sharp cut in ultraperipheral condition becomes larger with larger value of invariant mass, Fig.~\ref{fig:2}(b). However, applying the same type of the condition to the impact parameter, i.e. comparing the results from the SuperChic generator and our calculation, gives a difference of about $2\%$. A similar conclusion arises after comparing the total cross sections listed in Tab.~\ref{tab1}. Taking into account the possibility of different initial UPC condition, the used methods describe the experimental data of ATLAS \cite{ATLAS1,ATLAS2,ATLAS3} and CMS \cite{CMS1} in a similar way.

\begin{table}[!h]
	\begin{tabular}{c c c c c } 
		\hline
		\textbf{Experiment}        & $p_{t,min}$ [GeV] & UPC condition & $\sigma_{tot}$ [nb]& \\ \hline
		ATLAS   & 2.5  &    b $>$ 14 fm     & 81.062 $\pm$ 0.05 \\
		&      &    S$^2$(b) Eq.~(\ref{eq:s2b})       & 78.092 $\pm$ 0.05 \\
		&      &    SuperChick      & 76.421 $\pm$ 0.074 \\
		\hline
		CMS & 2 &   b $>$ 14 fm   & 105.986 $\pm$ 0.067 \\
		&   &   S$^2$(b)      & 102.104 $\pm$ 0.057 \\
		&   &   SuperChick    & 100.101 $\pm$ 0.144 \\ \hline
	\end{tabular}
	\caption{Total cross section for PbPb$\rightarrow$PbPb$\gamma\gamma$ in nb obtained in different approaches for experimental ATLAS/CMS kinematics: collision energy $\sqrt{s_{NN}}= 5.02$~TeV, di-photon invariant mass $M_{\gamma\gamma} > 5$~GeV, photon rapidity $|y|<2.4$. ATLAS and CMS have detected photons in different range of transverse momenta.}
	\label{tab1}
\end{table}

\begin{figure}[!h]
	(a)\includegraphics[scale=0.3]{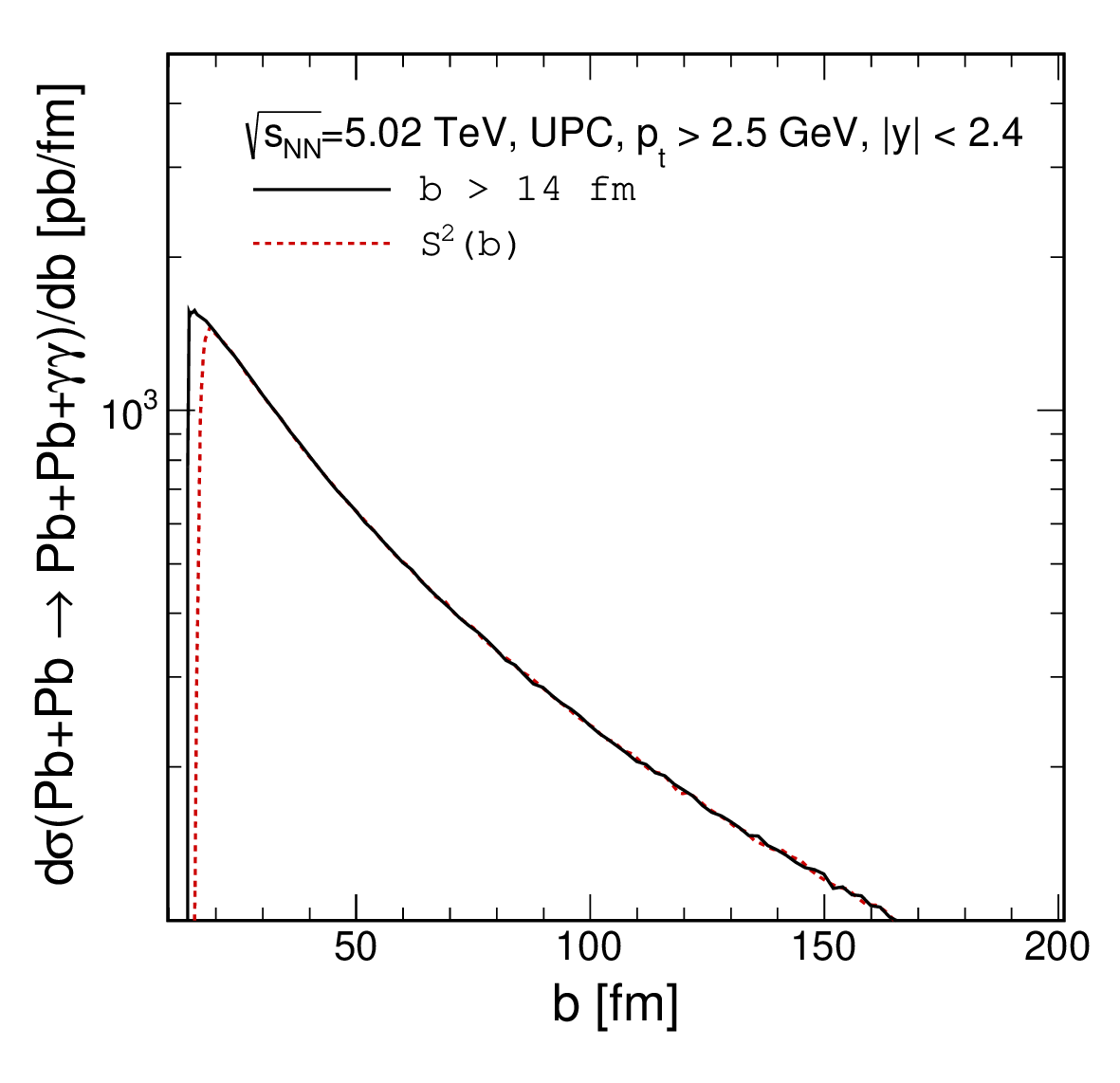}
	(b)\includegraphics[scale=0.3]{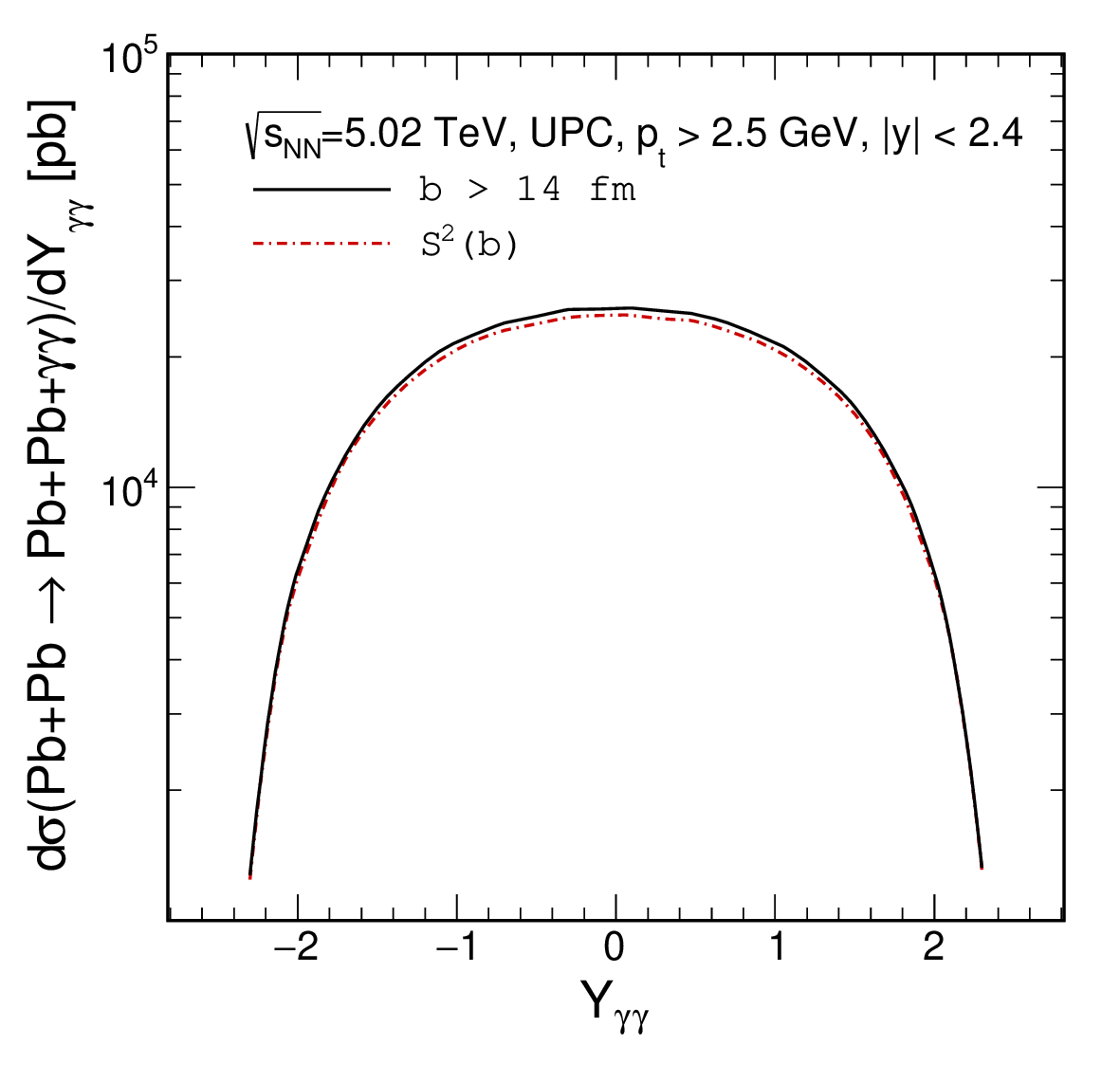}
	\caption{\label{fig:3} (a) Impact parameter and (b) rapidity distribution for UPC of lead-lead at collision energy $\sqrt{s_{NN}}=5.02$~TeV. The solid line corresponds to a sharp cut on the impact parameter and the red dash-dotted line is for the absorption factor as given by Eq.~(\ref{eq:s2b}). }
\end{figure}

 In Fig.~\ref{fig:3} we show
results with a sharp cut-off, $ b >14$~fm, and when including the smooth dependence
of $S^2_{abs}(b)$ on impact parameter. In Fig.~\ref{fig:3}(a), we show somewhat academic impact 
parameter distribution (not measureable) while in Fig.~\ref{fig:3}(b), rapidity distribution of outgoing system, $Y_{\gamma\gamma} = (y_1 + y_2)/2$. 
Only small differences due to the treatment in the $b$-space (about $(3-4)\%$)
at midrapidities can be observed.

Having described the ATLAS data, we wish to discuss new unexplored kinematics regions.

\subsubsection{Broad range of rapidity, full phase space}

\begin{figure}[!h]
	\includegraphics[scale=0.4]{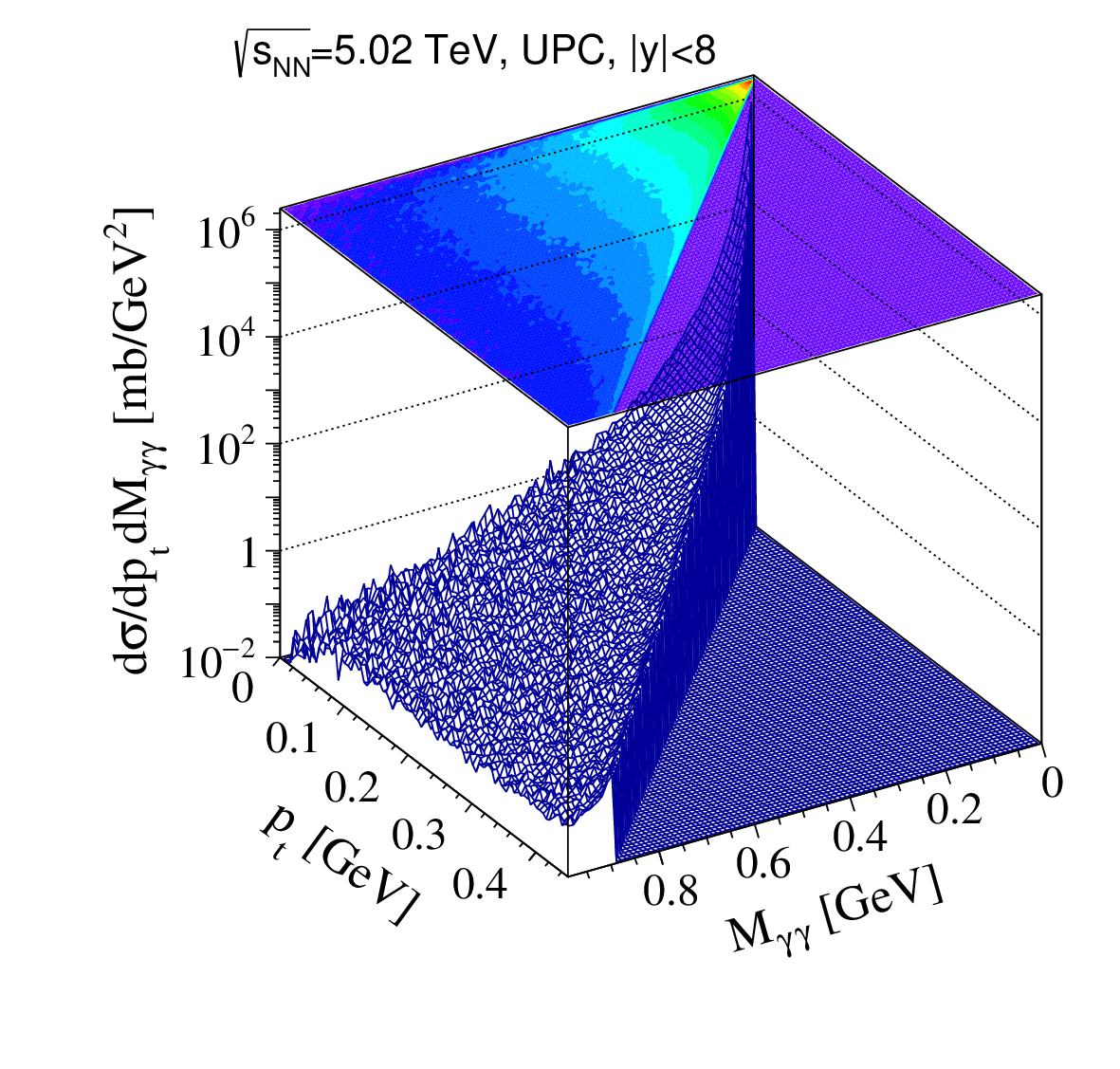}
	\caption{\label{fig:4} Two-dimensional distribution in photon transverse momentum and di-photon invariant mass for UPC of lead-lead in a wide photon rapidity range $y\in(-8,8)$, starting from $p_t> 5$~MeV.}
\end{figure}

Now we go to the somewhat broader range of rapidity and allow for very
small transverse momenta. In Fig.\ref{fig:4} we show
two-dimensional distribution in $(M_{\gamma \gamma}, p_{t,\gamma})$. 
We observe a strong enhancement for $M_{\gamma \gamma} \approx 2 p_t$, which infers that small $p_{t,\gamma}$ means automatically small $M_{\gamma\gamma}$ and vice versa.

\begin{table}[!h]
	\begin{tabular}{c c c} 
		\hline
		& $\sigma_{tot}$ [mb]& \\ \hline
		Total      & 91.675 $\pm$ 0.023 \\ 
		electrons+muons        & 41.597 $\pm$ 0.010 \\
		electrons  & 39.163 $\pm$ 0.010 \\
		quarks     & 12.483 $\pm$ 0.003 \\ \hline
	\end{tabular}
	\caption{Total cross section in mb for PbPb$\rightarrow$PbPb$\gamma\gamma$ for different fermionic contributions artificially separated. Here collision energy $\sqrt{s_{NN}}= 5.02$~TeV, di-photon invariant mass M$_{\gamma\gamma} = (0.01 - 1)$~GeV, photon transverse momentum p$_t>5$~MeV and photon rapidity $|y|<8$.}
	\label{tab2}
\end{table}

\begin{figure}[!h]
	(a)\includegraphics[scale=0.32]{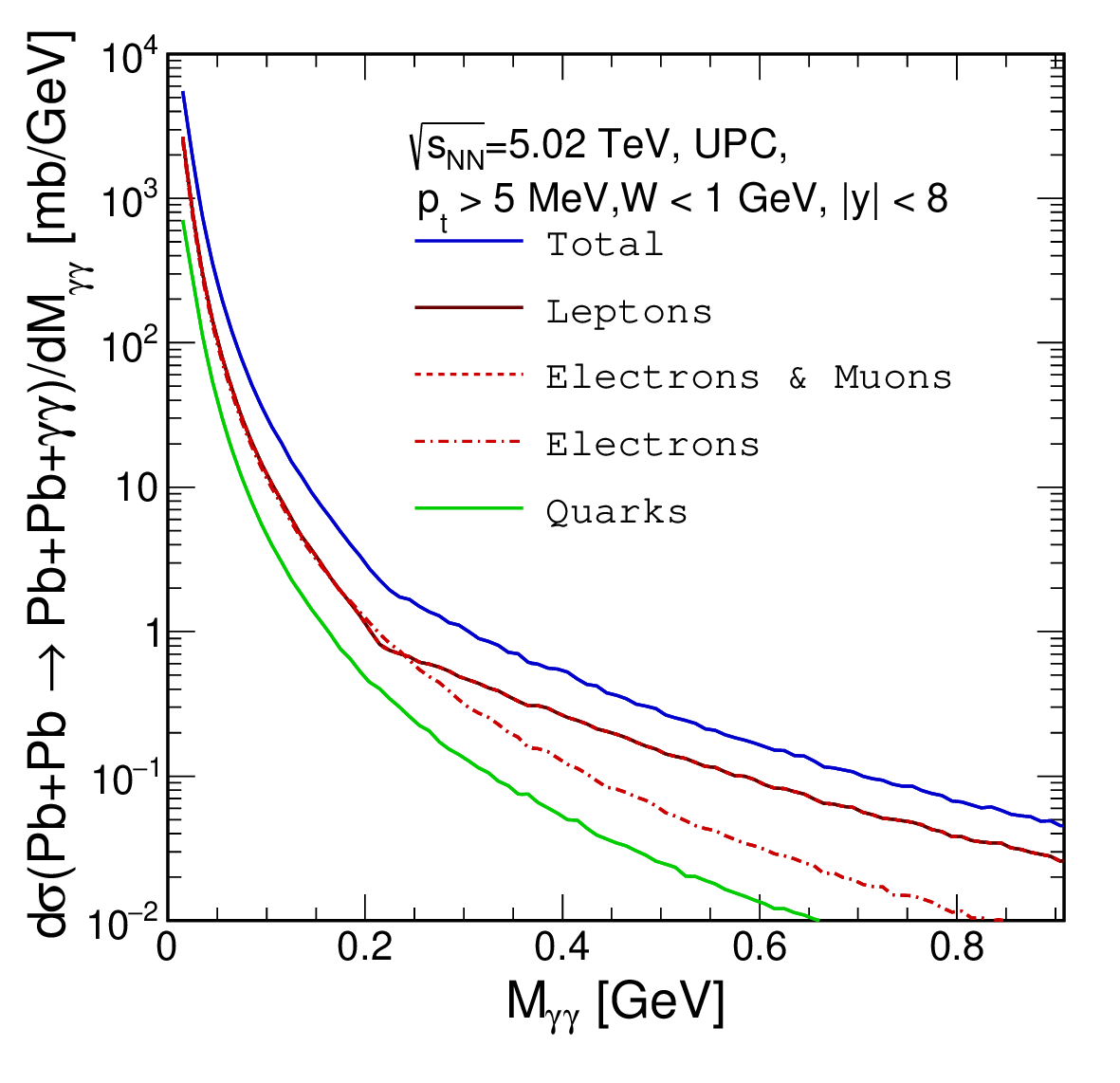}
	(b)\includegraphics[scale=0.32]{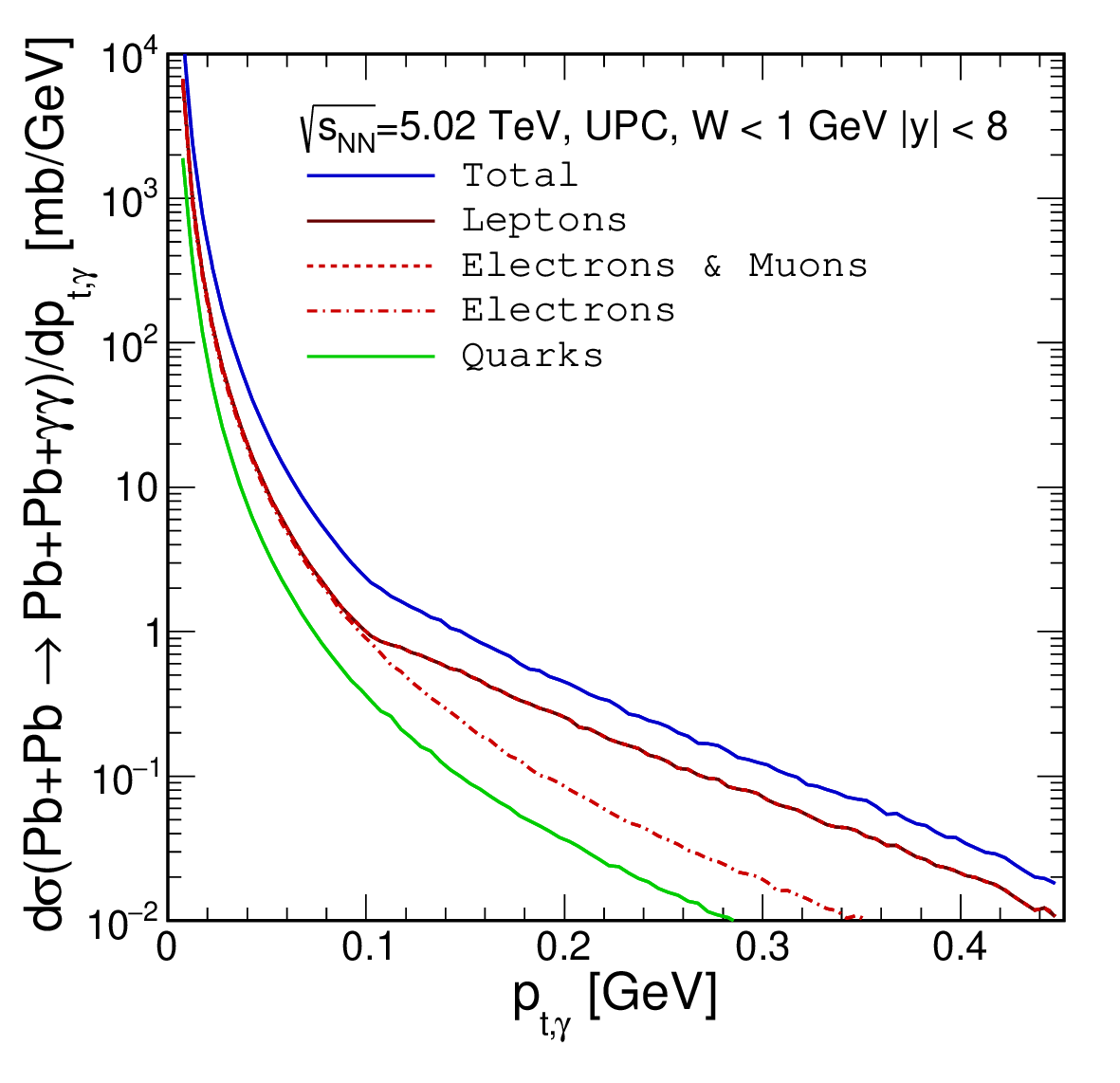}
	(c)\includegraphics[scale=0.32]{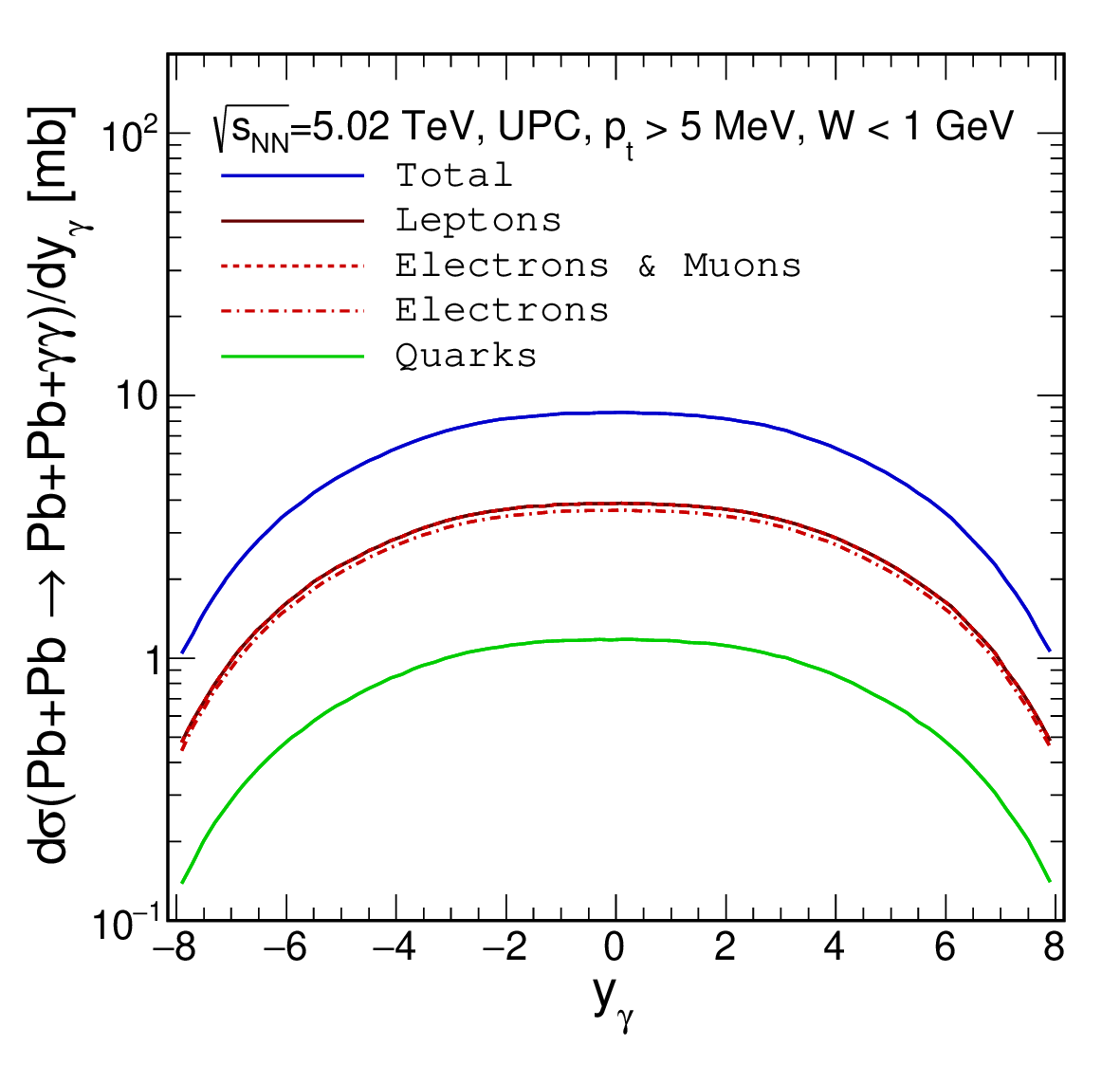}
	\caption{\label{fig:5} (a) Di-photon invariant mass, (b) photon transverse momentum and (c) rapidity distribution in mb for different fermionic contributions in ultraperipheral lead-lead collisions at energy $\sqrt{s_{NN}}=5.02$~TeV. The blue solid line corresponds to a sum of all contributions, the red solid line is a sum for leptons, the dashed red line relates to electrons, and the green solid line represents quarks contribution.}
\end{figure}

In Fig.\ref{fig:5} we show different distributions
in (a) $M_{\gamma \gamma}$, (b) $p_t$ and in (c) $y_\gamma$. In this figure
we present decomposition into different loop contributions (leptons, quarks).
The electron/positron loops dominate at low $M_{\gamma \gamma}$ and
low $p_t$. Despite the noticeable difference between leptonic and quarkonic contribution, their coherent sum contributes much more than the leptonic contribution alone. Looking at the total cross section values in Tab.~\ref{tab2}, although the quark contribution is more than twice as small as the lepton contribution ($\sigma_{tot}^{quarks} \approx 30 \% \, \sigma_{tot}^{leptons}$), summing over the helicities for leptonic and quarkish boxes, yields a result that is more than twice bigger than the lepton contribution ($\sigma_{tot}^{leptons} \approx 45 \% \, \sigma_{tot}^{fermions}$). These cross sections are calculated for the di-photon invariant mass range up to $1$~GeV.

\begin{figure}
	(a)\includegraphics[scale=0.32]{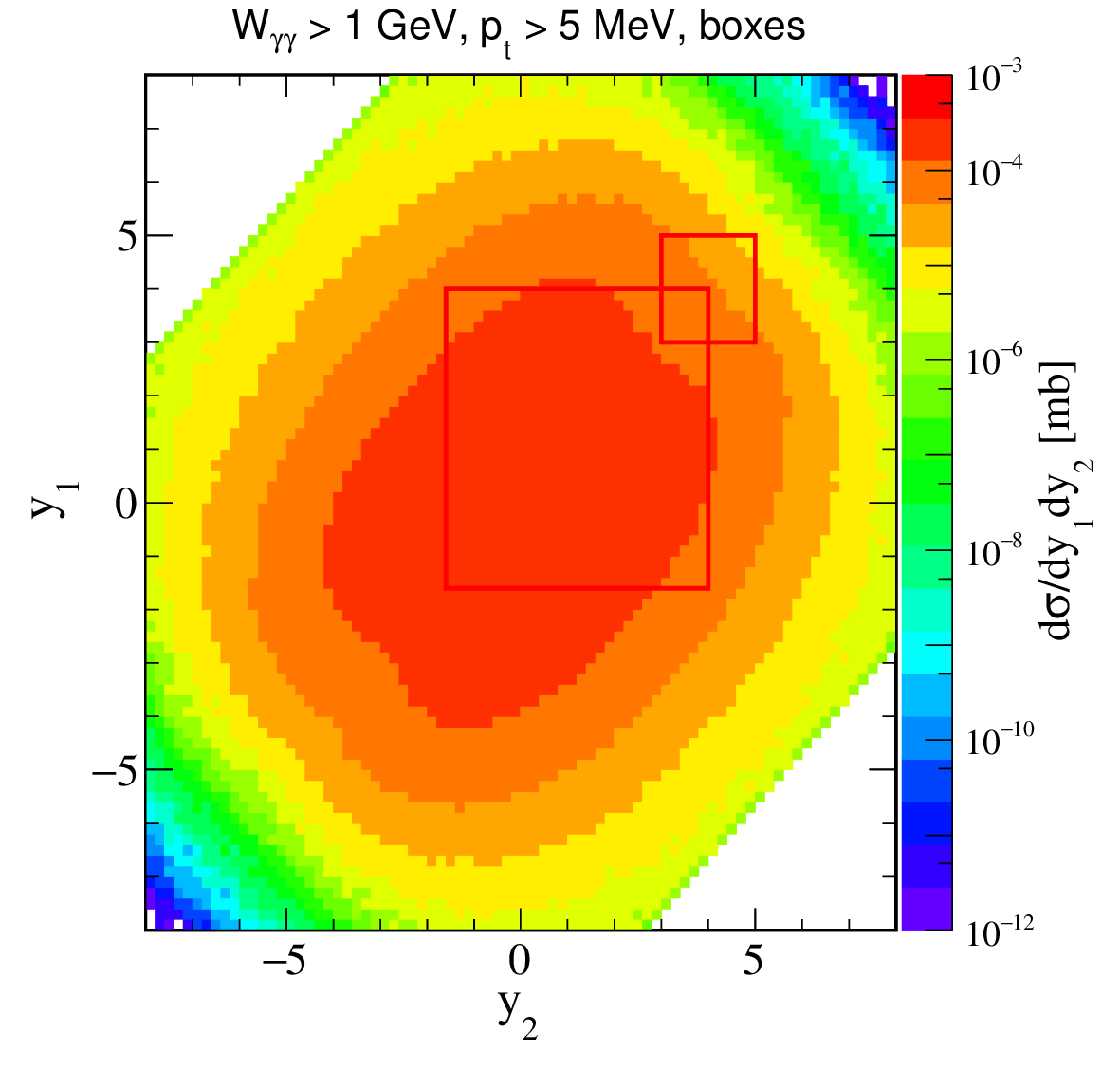}
	(b)\includegraphics[scale=0.32]{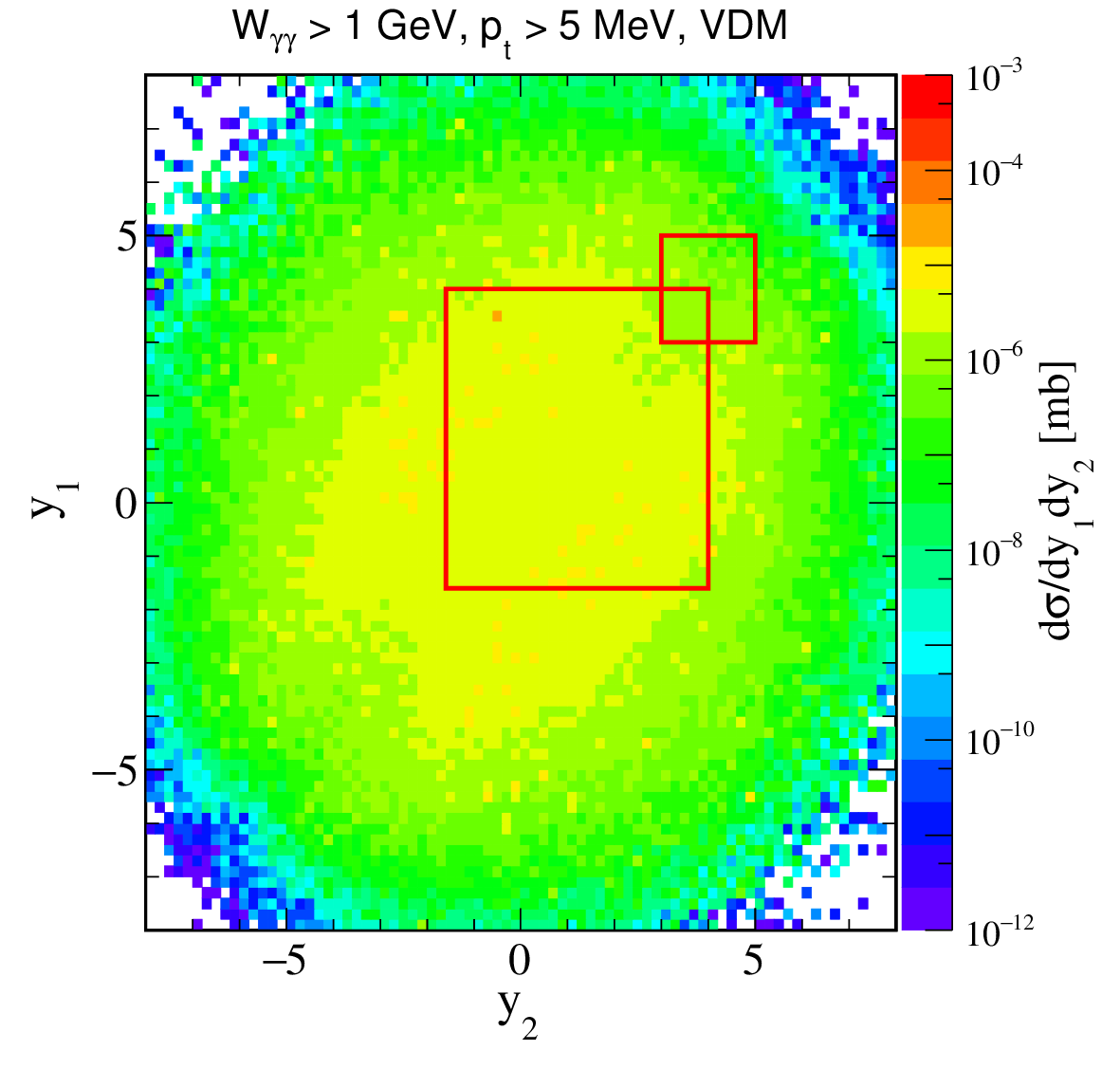}
	(c)\includegraphics[scale=0.32]{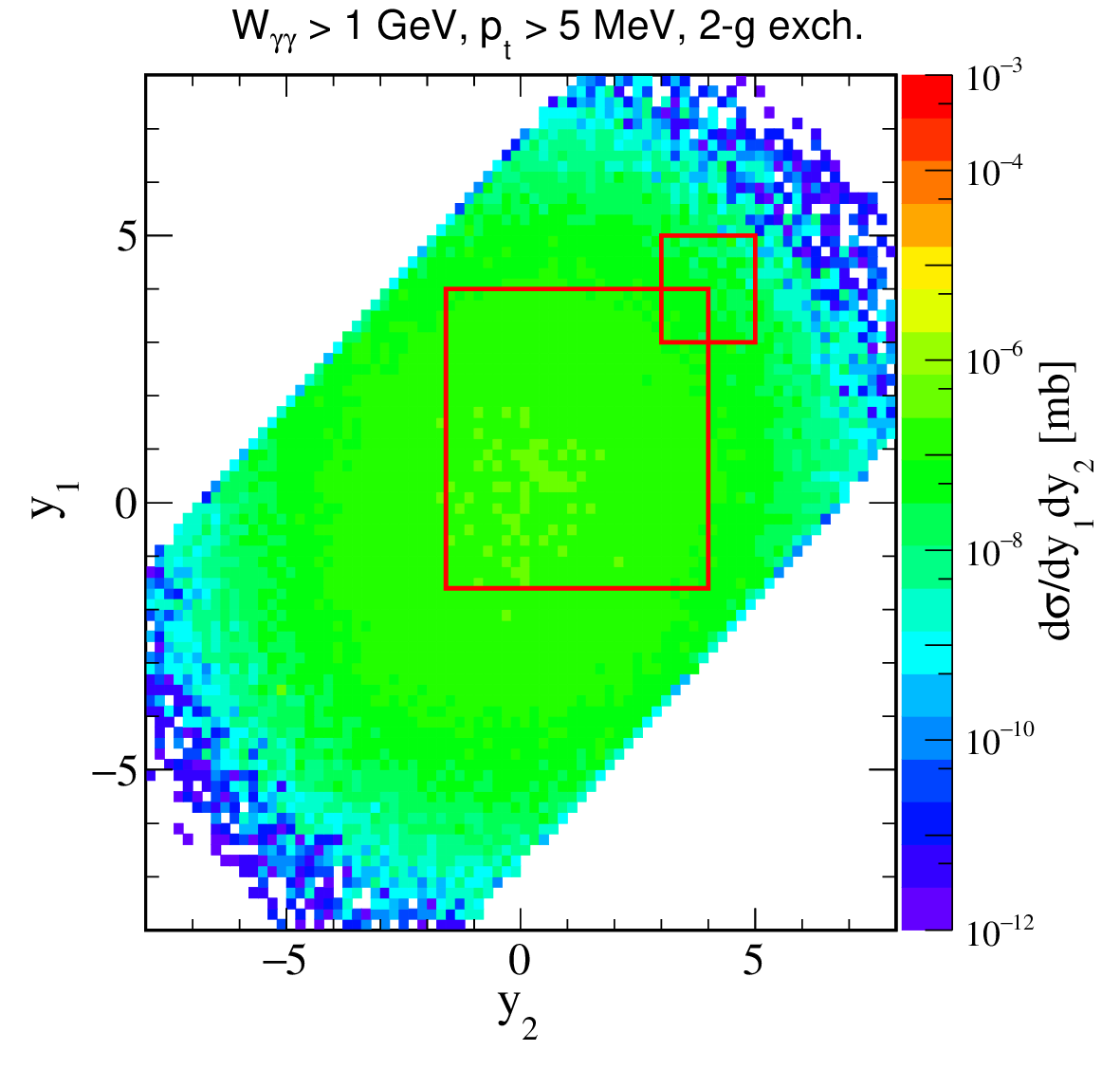}
	\caption{\label{fig:8} Distribution in ($y_1$,$y_2$) in mb for photon transverse momentum $p_{t}>5$~MeV, di-photon invariant mass $M_{\gamma\gamma}>1$~GeV. (a) Boxes, (b) VDM-Regge and (c) two-gluon exchange.}
\end{figure}

\begin{figure}
	(a)\includegraphics[scale=0.32]{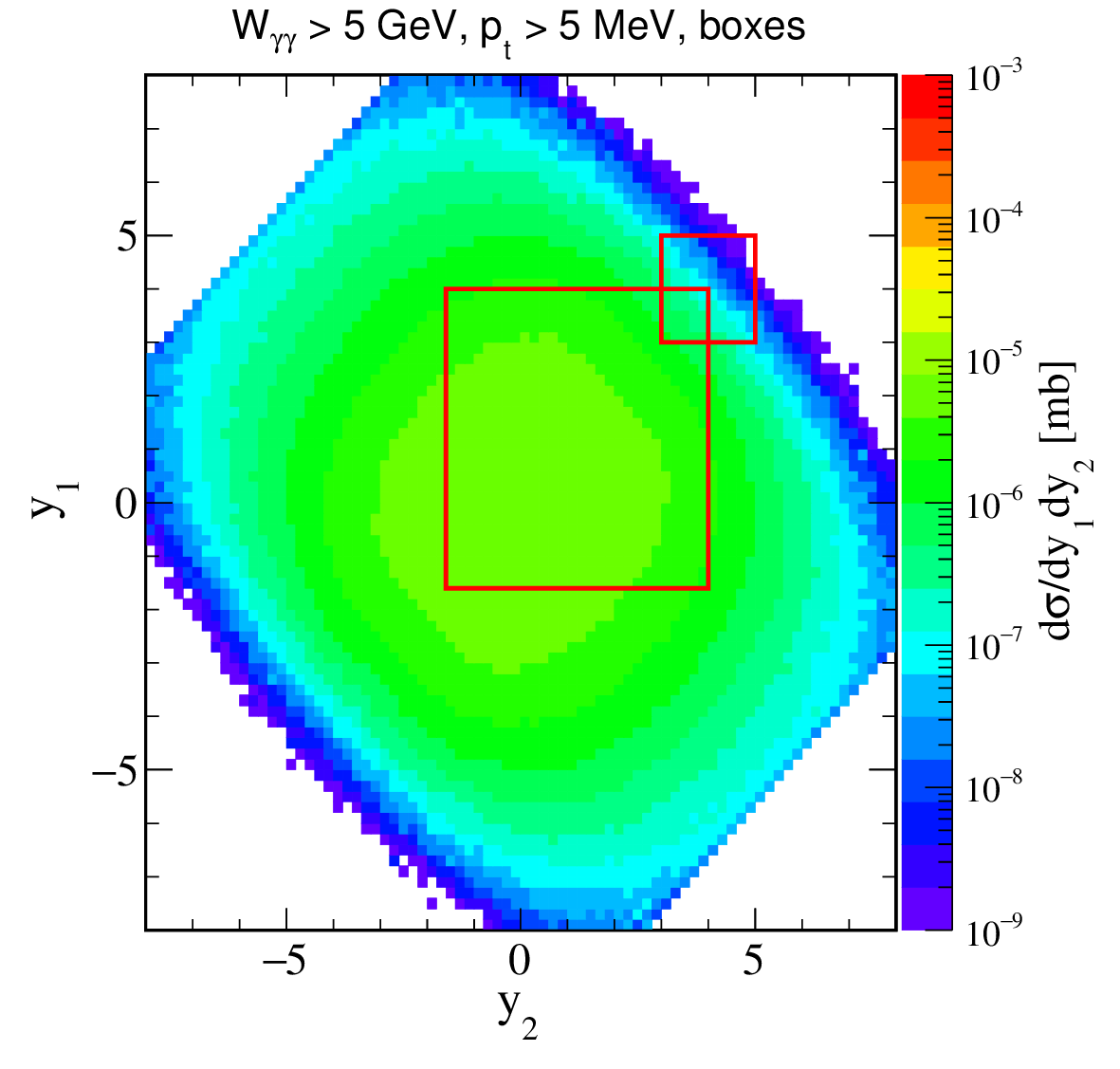}
	(b)\includegraphics[scale=0.32]{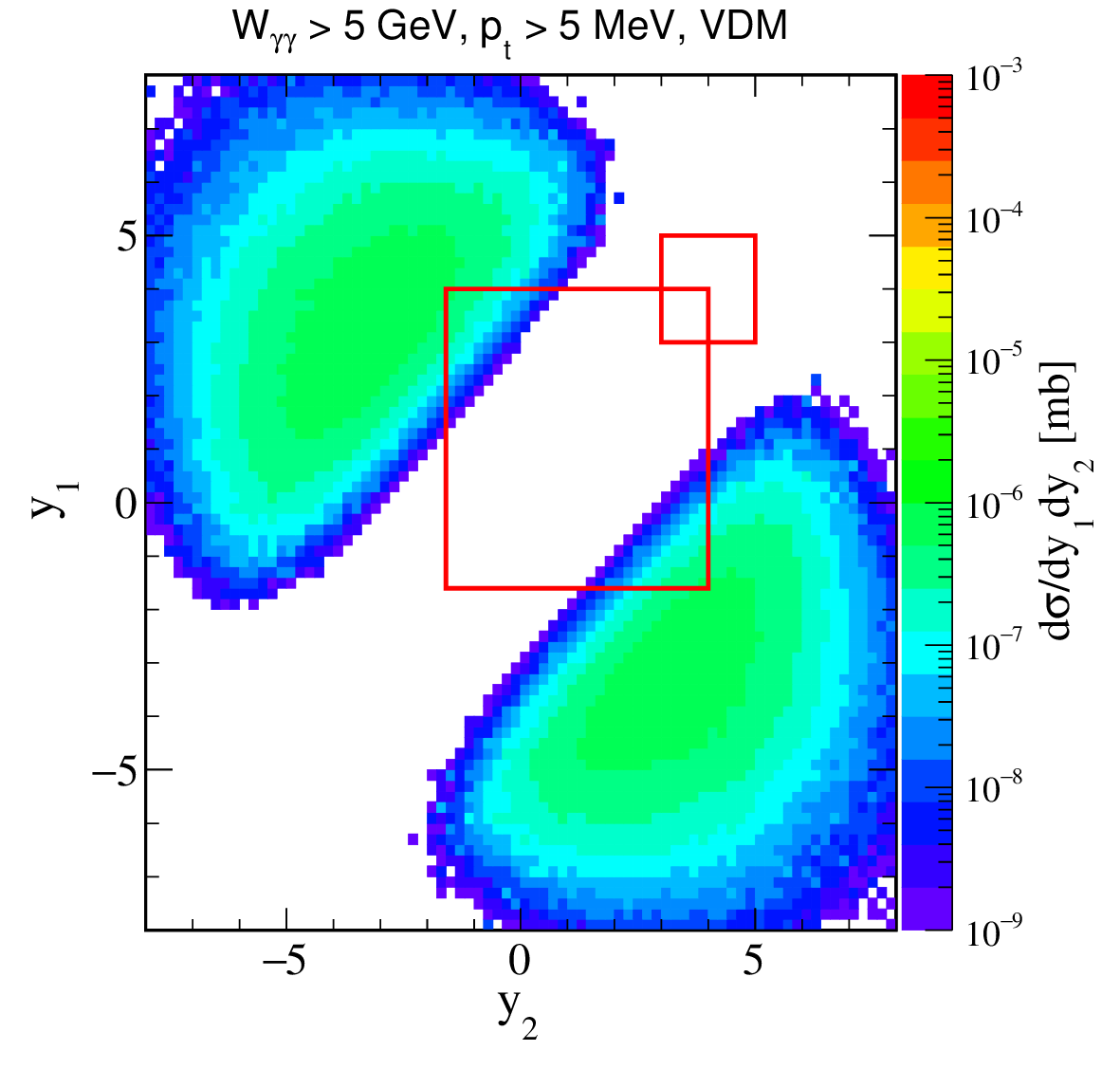}
	\caption{\label{fig:7.2} Distribution in ($y_1$,$y_2$) in mb for transverse momentum $p_{t}>5$~MeV, di-photon invariant mass $M_{\gamma\gamma}>5$~GeV. (a) Boxes, (b) VDM-Regge mechanism.}
\end{figure}

In Fig.~\ref{fig:8} and Fig.~\ref{fig:7.2} we show corresponding two-dimensional
distributions ($y_1$, $y_2$) for three different mechanisms separately:
boxes, double-hadronic fluctuations and two-gluon exchange.
In this inclusive case (integration over $p_t = p_{1t} = p_{2t}$ and
$W_{\gamma \gamma}$)  we observe absolute dominance of the box
contribution over the other mechanisms. The red contours represent the rapidity limit for ALICE 3: $-1.6<y_\gamma<4$ and $3<y_\gamma<5$. The largest cross section occurs in the midrapidity of both photons. Within this rapidity range, it is expected to be able to detect photons with $p_{t,\gamma}>100$~MeV. Our predictions show that in the forward range of the calorimeter, $3<y_\gamma<5$, even though the VDM-Regge contribution is about three orders of magnitude smaller than fermionic boxes, it should be included to determine the coherent sum of these two processes.

\begin{figure}[!h]
	(a)\includegraphics[scale=0.32]{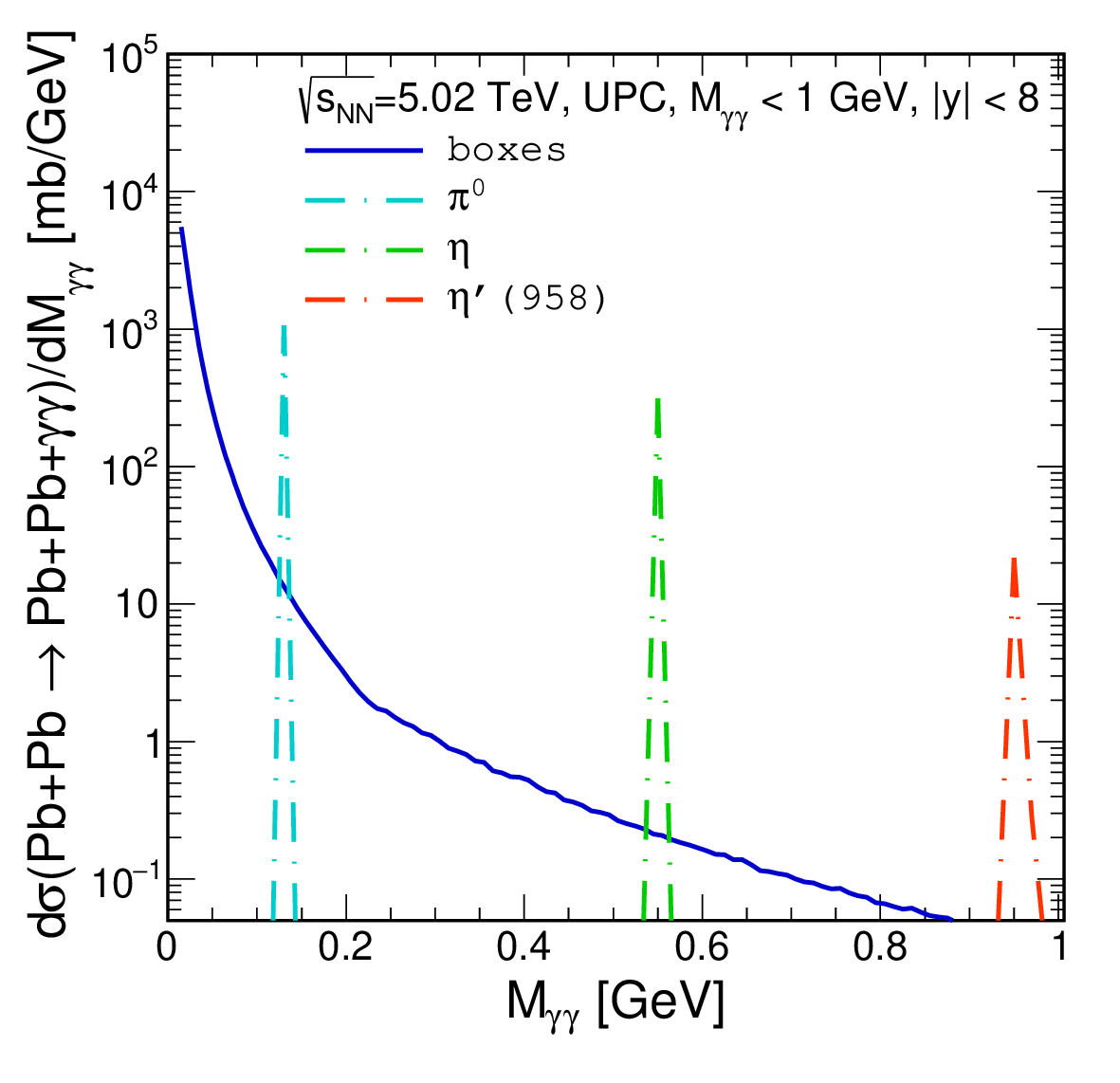}
	(b)\includegraphics[scale=0.32]{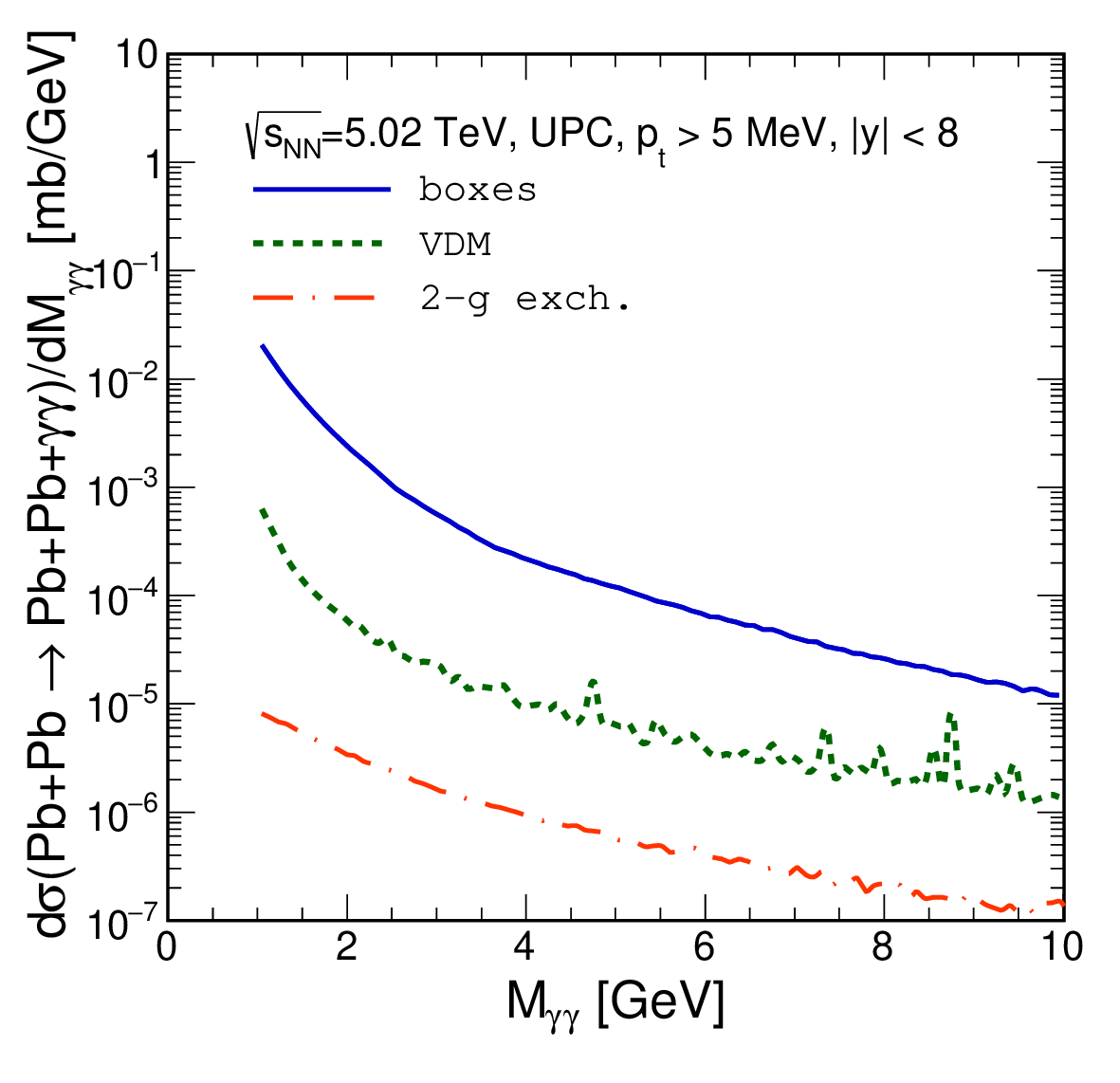}
	\caption{\label{fig:9} Di-photon invariant mass distribution for photon transverse momentum $p_{t}>5$~MeV and photon rapidities $y_{1/2} \in (-8,8)$. (a) Boxes vs resonances, $M_{\gamma\gamma}<1$~GeV (b) Boxes vs VDM-Regge and vs two-gluon exchange contribution, $M_{\gamma\gamma}>1$~GeV.}
\end{figure}

In Fig.\ref{fig:9} we present di-photon invariant mass
distributions for different mechanisms including (a) $\pi^0$, $\eta$ and
$\eta'$ resonances as well as (b) VDM-Regge and two-gluon exchange mechanism. While the resonances stick over the box-continuum
they can be easily eliminated imposing windows around the resonance
positions. Although at first glance the contribution from higher order mechanisms seems insignificant, it is important to be aware that the contribution from VDM-Regge is only an order of magnitude smaller than the leading contribution from the fermionic boxes.

\begin{figure}
	(a)\includegraphics[scale=0.32]{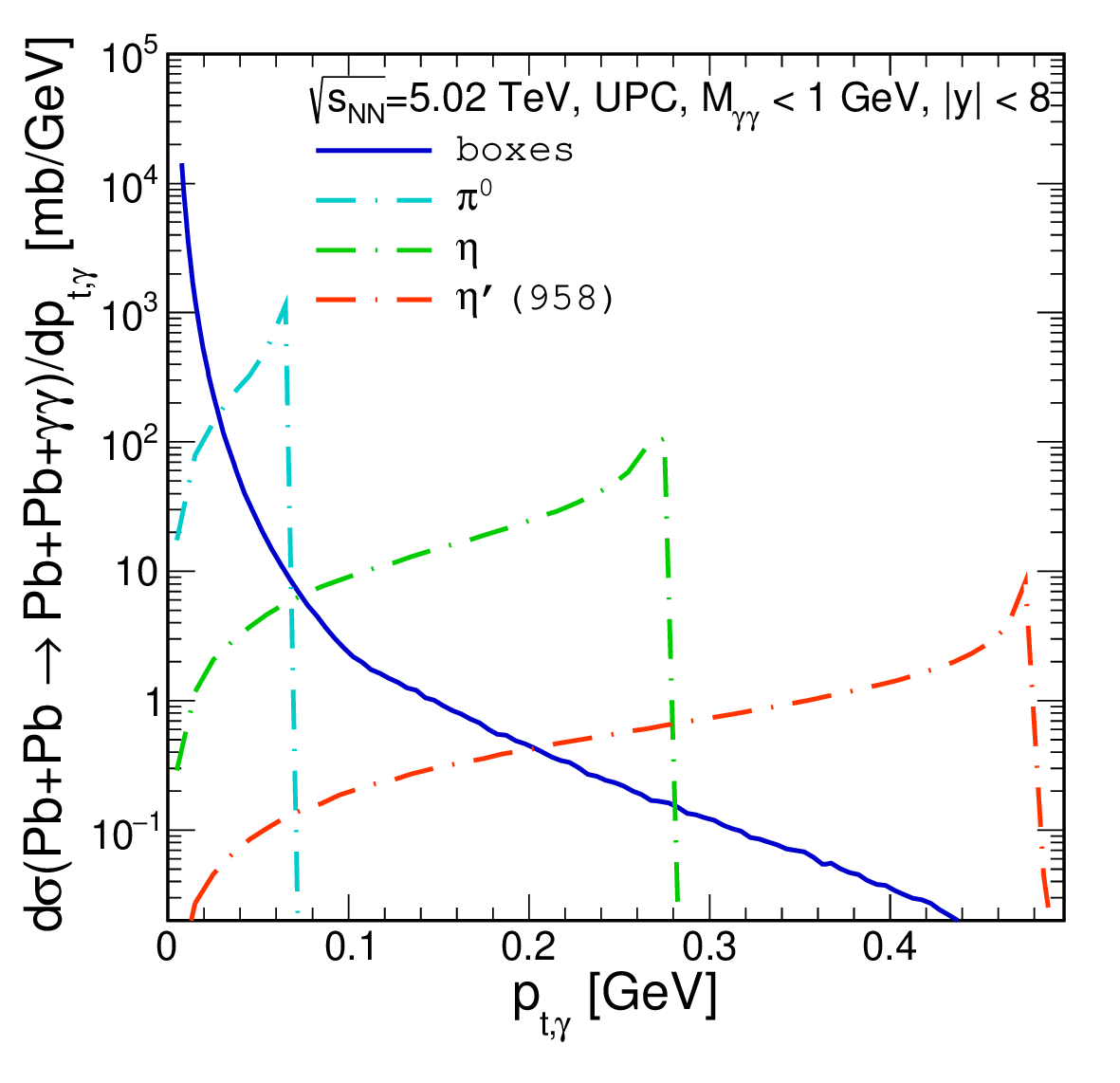}
	(b)\includegraphics[scale=0.32]{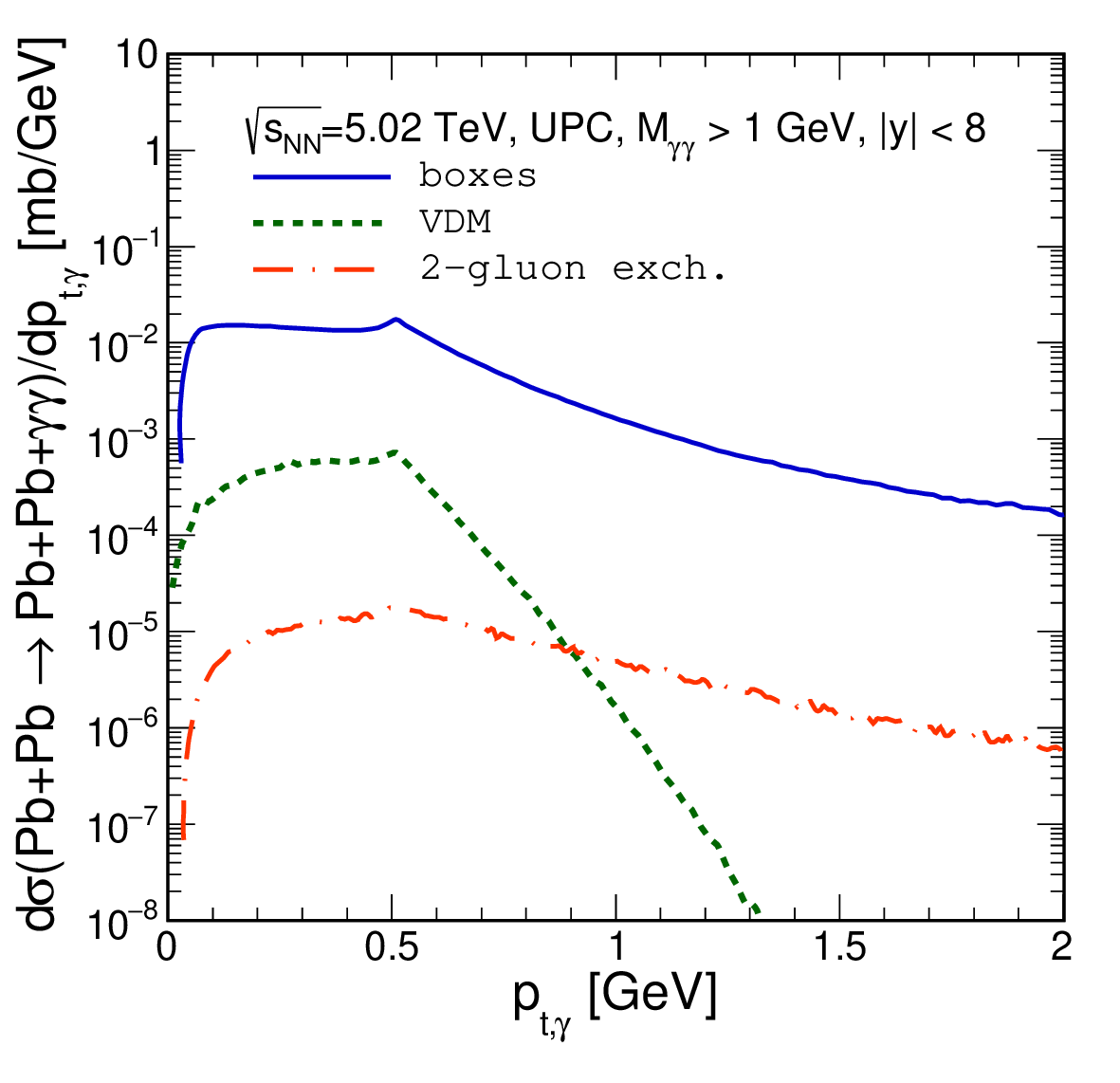}
	\caption{\label{fig:10} Transverse momentum distribution for photon transverse momentum $p_{t}>5$~MeV, di-photon invariant mass $W_{\gamma\gamma}<1$~GeV and photon rapidities $y_{1/2} \in (-8,8)$. (a) Boxes vs resonances; (b) boxes vs VDM-Regge vs two-gluon exchange.}
\end{figure}

The corresponding distributions in $p_t = p_{1t} = p_{2t}$ are shown
in Fig.\ref{fig:10}. We observe huge enhancements of the cross
section at $p_t \sim M_R/2$ (jacobian peak). It would be interesting 
to see such enhancements experimentally for controlling the general situation. 
Imposing the windows around resonances would allow to eliminate 
the resonance contributions. However, this would probably distort to some extent 
other distributions, in particular, those for 
$p_t = p_{1t} = p_{2t}$.
Therefore it is not clear to us whether such cuts would be welcomed.

\begin{figure}[!h]
	(a)\includegraphics[scale=0.32]{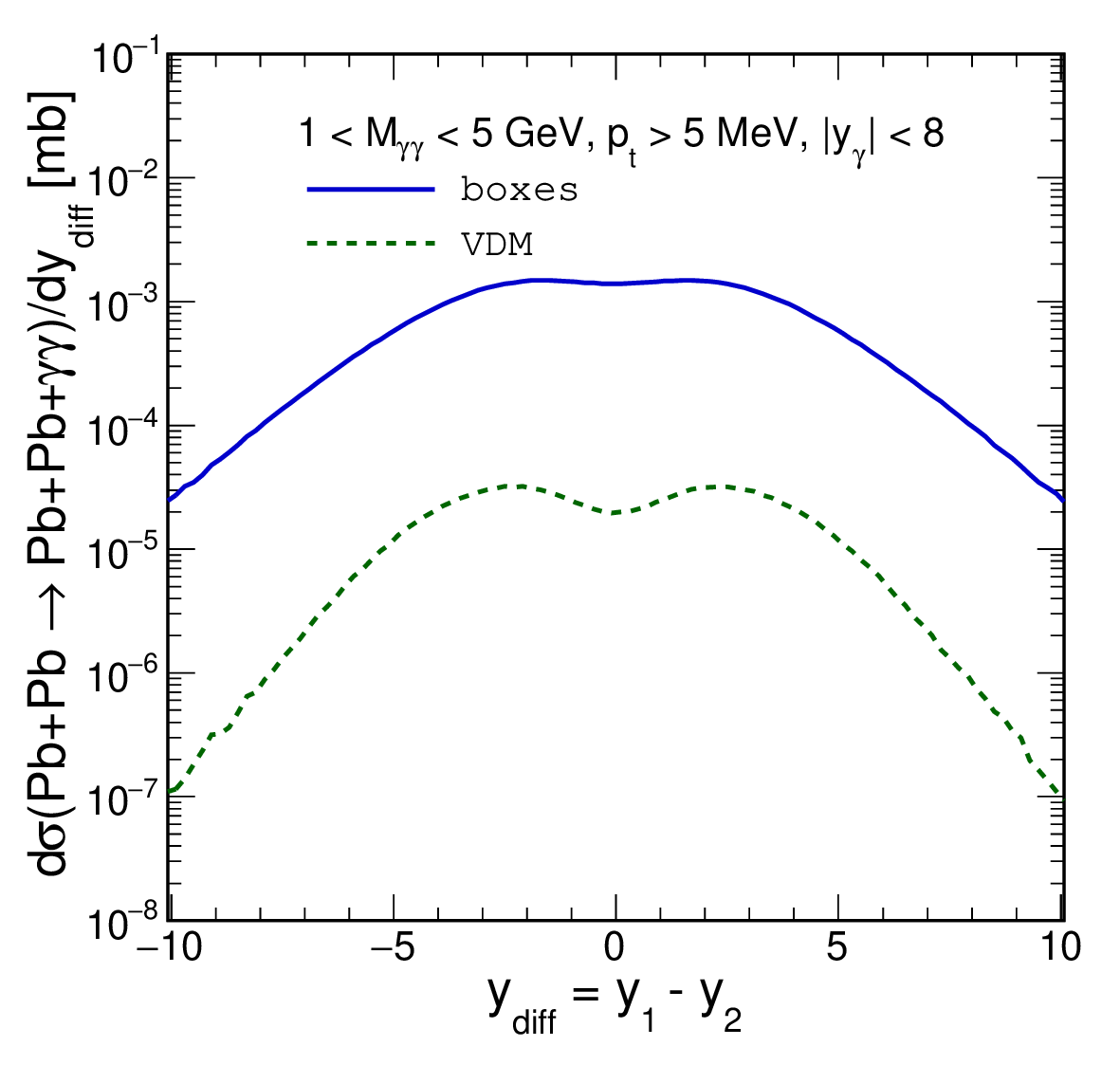}
	(b)\includegraphics[scale=0.32]{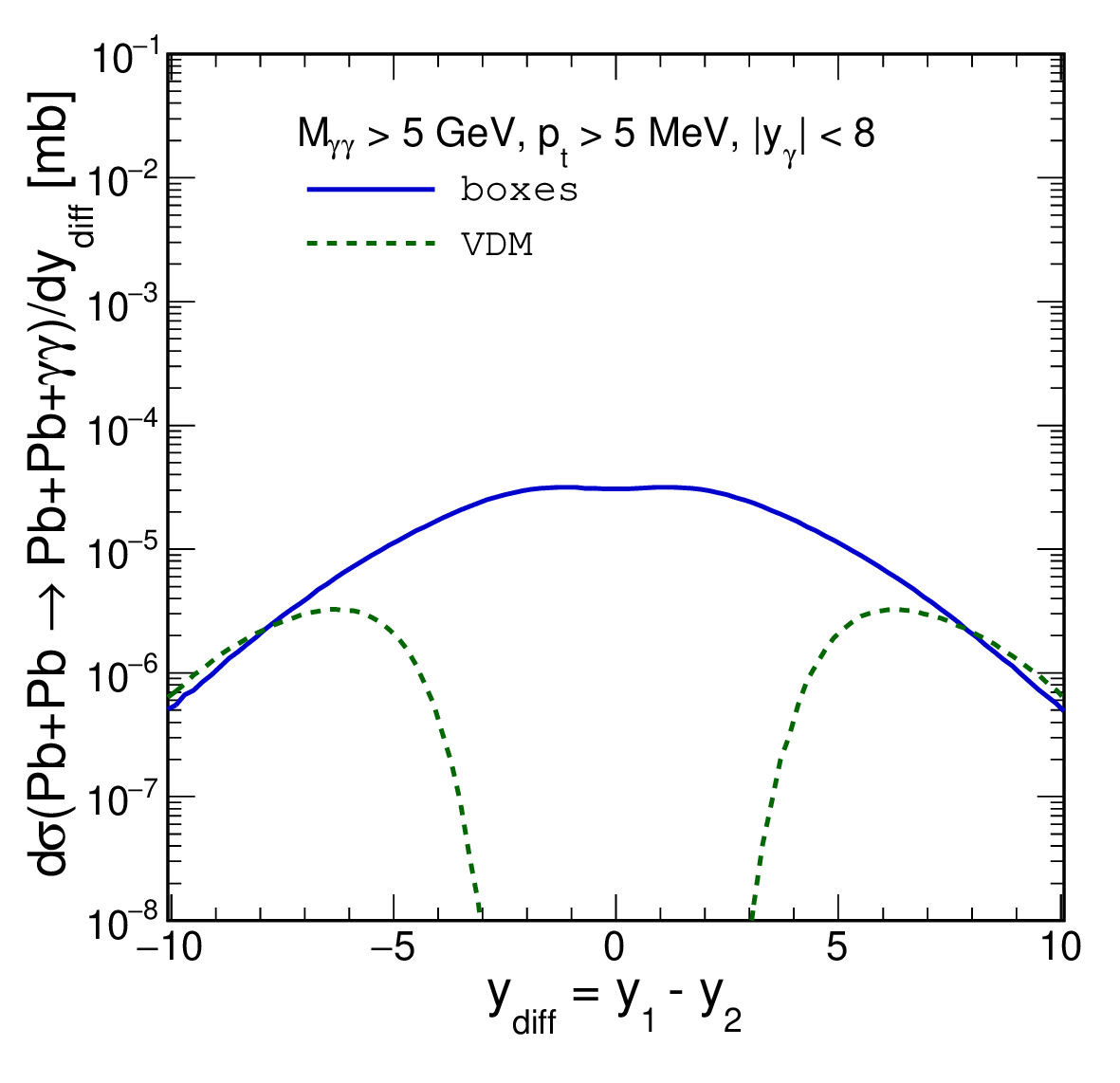}
	\caption{\label{fig:11} Distribution in $y_{diff}$ for light-by-light scattering processes in PbPb$\to$PbPb$\gamma\gamma$. Here the transverse momentum cut is equal to 5~MeV. The blue solid line relates to the boxes, and the green dotted line to the VDM-Regge contribution. Here the range of measured di-photon invariant mass is: (a) $(1-5)$~GeV, (b) $> 5$~GeV.}
\end{figure}

For completeness, in Fig.\ref{fig:11} we show distributions
in rapidity difference hoping it could distinguish different
mechanisms. Indeed the shape of the distributions corresponding to
two-photon hadronic fluctuations seems somewhat broader than that for boxes. There is a region
of $y_{diff}$ where the VDM-Regge contribution is as big as that for fermionic boxes.
This region of phase space is, however, not easy to measure at the LHC.
The resonance contributions, not shown here explicitly, are concentrated
at $-2<y_{diff}<2$ but for $M_{\gamma\gamma}>5$~GeV the light mesons are automatically removed. We shall discuss the resonance contribution for ALICE 3.

\subsubsection{ALICE and FoCal}

The FoCal detector planned for Run 4 was described in \cite{FoCAL}. It is a general purpose detector. It can also measure photons.

\begin{figure}[!h]
	(a)\includegraphics[scale=0.32]{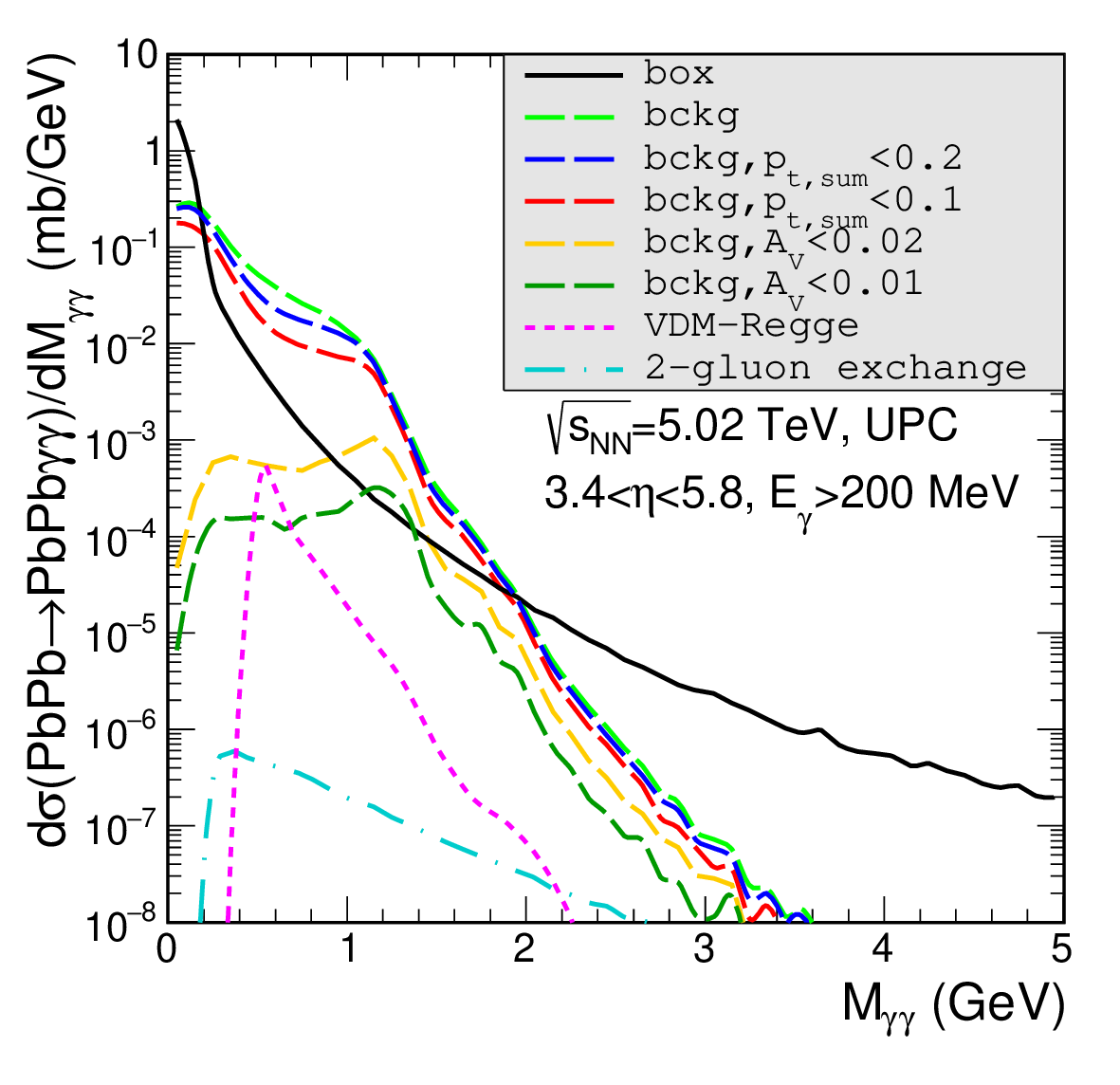}
	(b)\includegraphics[scale=0.32]{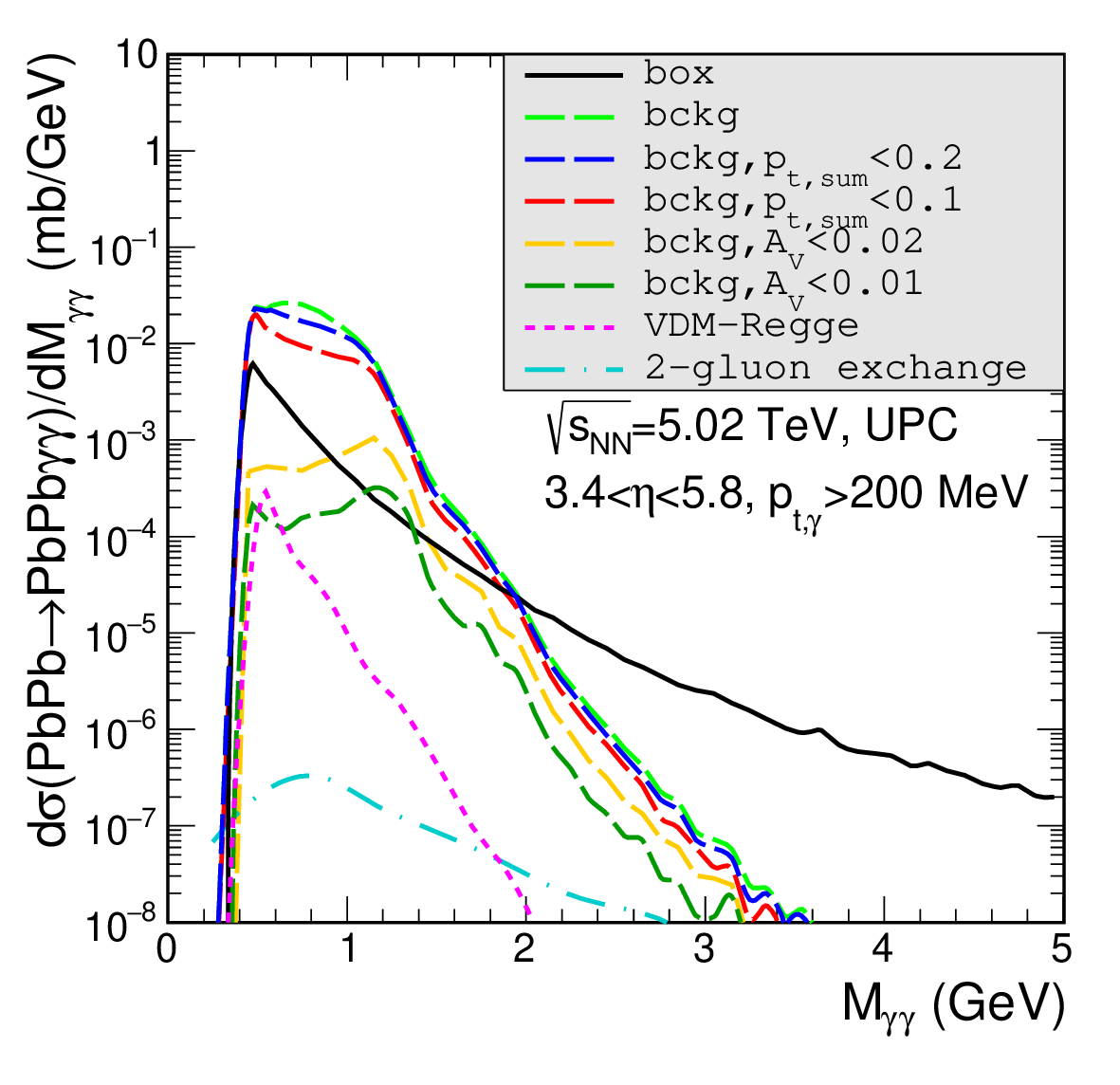}
	\caption{\label{fig:dsig_dW_ALICE_FoCal_2} Invariant mass distribution for the nuclear process. Predictions are made for the future FoCal acceptance, i.e. (a) $E_{t,\gamma}>200$~MeV and $3.4<y_{\gamma_{1/2}}<5.8$, (b) $p_{t,\gamma}>200$~MeV and $3.4<y_{\gamma_{1/2}}<5.8$. Here, both photons are ''measured'' in FoCal. The background contribution is presented for different elimination cuts.}
\end{figure}

We start our presentation from the results when in addition both photons are measured by FoCal. In Fig.\ref{fig:dsig_dW_ALICE_FoCal_2} (a) we show results when both photons have energies bigger than
200~MeV. In addition, we show the $\pi^0\pi^0$ background. In this case, only two photons are measured. Without additional cuts, the background is clearly bigger than the signal. However, by imposing extra conditions on vector asymmetry, we can lower the background contribution.
The vector asymmetry defined in Ref.~\cite{KSS} as:
\begin{equation}
A_V = |\vec{p}_{t,1}-\vec{p}_{t,2}|/|\vec{p}_{t,1}+\vec{p}_{t,2}| \;,
\end{equation} 
reflects a convolution of each photon transverse momentum vector.

At very low $M_{\gamma\gamma}$ the $\pi^0\pi^0$ background is negligible, which opens a new window to measure the $\gamma\gamma\to\gamma\gamma$ scattering at $M_{\gamma\gamma}<1$~GeV while for the ATLAS experiment it was $M_{\gamma\gamma}>5$~GeV.
In Fig.\ref{fig:dsig_dW_ALICE_FoCal_2}(b) we show similar results when imposing $p_{t,\gamma}$ condition. In this case, it is rather difficult to eliminate the $\pi^0\pi^0$ background. Here the VDM-Regge component is relatively small. Only at $W_{\gamma\gamma} \approx 5$~GeV some effect of the VDM-Regge component could potentially be observed.
Comparing the results for the same cut on the energy of each outgoing photon, Fig.~\ref{fig:dsig_dW_ALICE_FoCal_2}(a), and on the transverse momentum of the photon, Fig.~\ref{fig:dsig_dW_ALICE_FoCal_2}(b), one can observe that the limit on $p_{t,\gamma}$ removes a very large contribution to the total cross section, which is located in a small di-photon invariant mass.

\begin{figure}[!h]
	\includegraphics[scale=0.32]{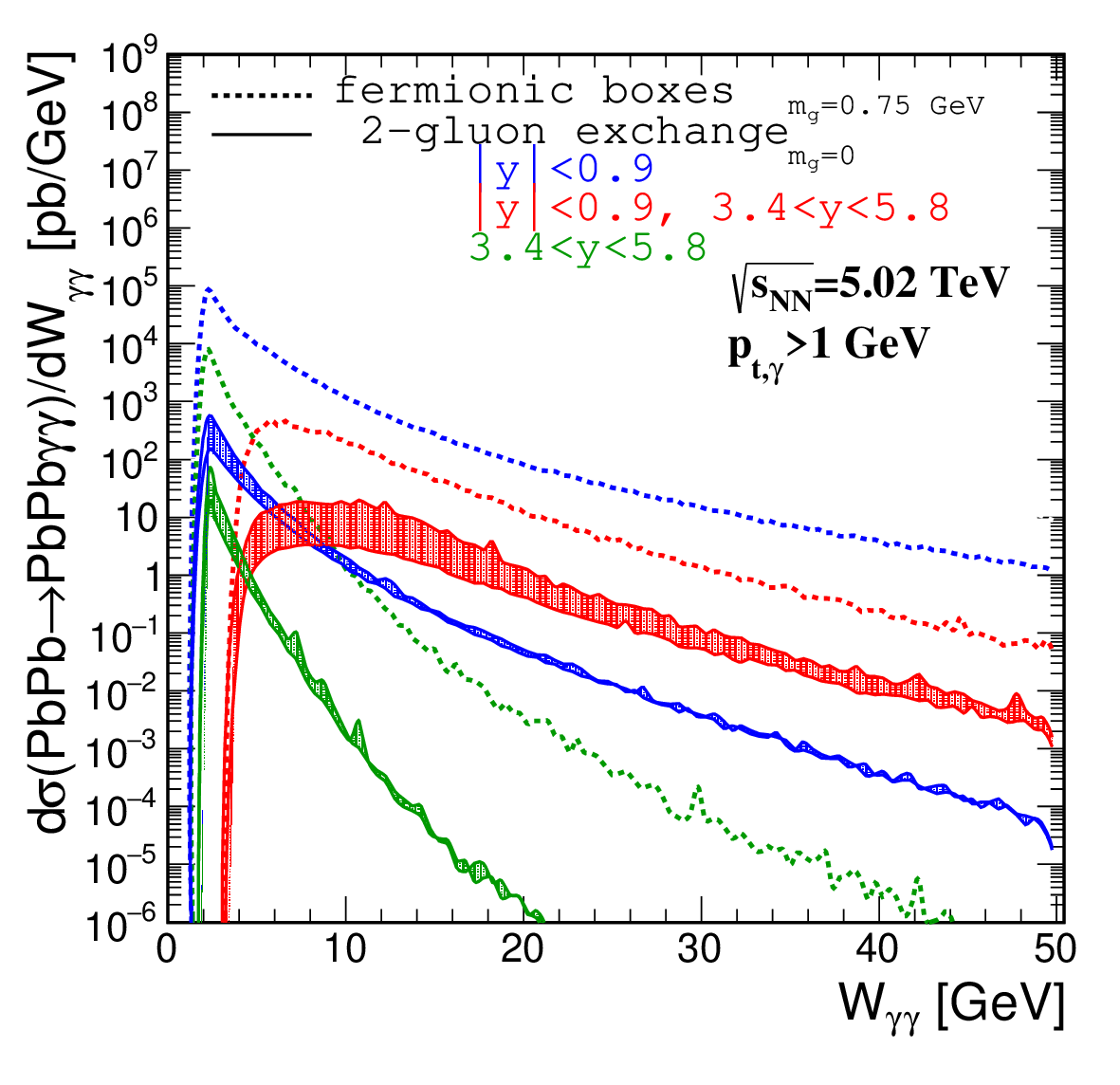}
	\caption{\label{fig:dsig_dW_ALICE_FoCal} Di-photon invariant mass distribution for PbPb$\to$PbPb$\gamma\gamma$ process for ALICE, FoCal and their combination.}
\end{figure}

Another option is to use simultaneously the FoCal and the main ALICE detector. In Fig.\ref{fig:dsig_dW_ALICE_FoCal}, we simultaneously show the box and two-gluon exchange contributions. In this calculation, we assumed that the transverse momenta of both photons have $p_t$~$>$~1~GeV. In this case, the separated two-gluon exchange contribution is only an order of magnitude smaller than the box contribution. The shaded band is due to ''unknown'' effective gluon mass.

\begin{figure}[!h]
	(a)\includegraphics[scale=0.3]{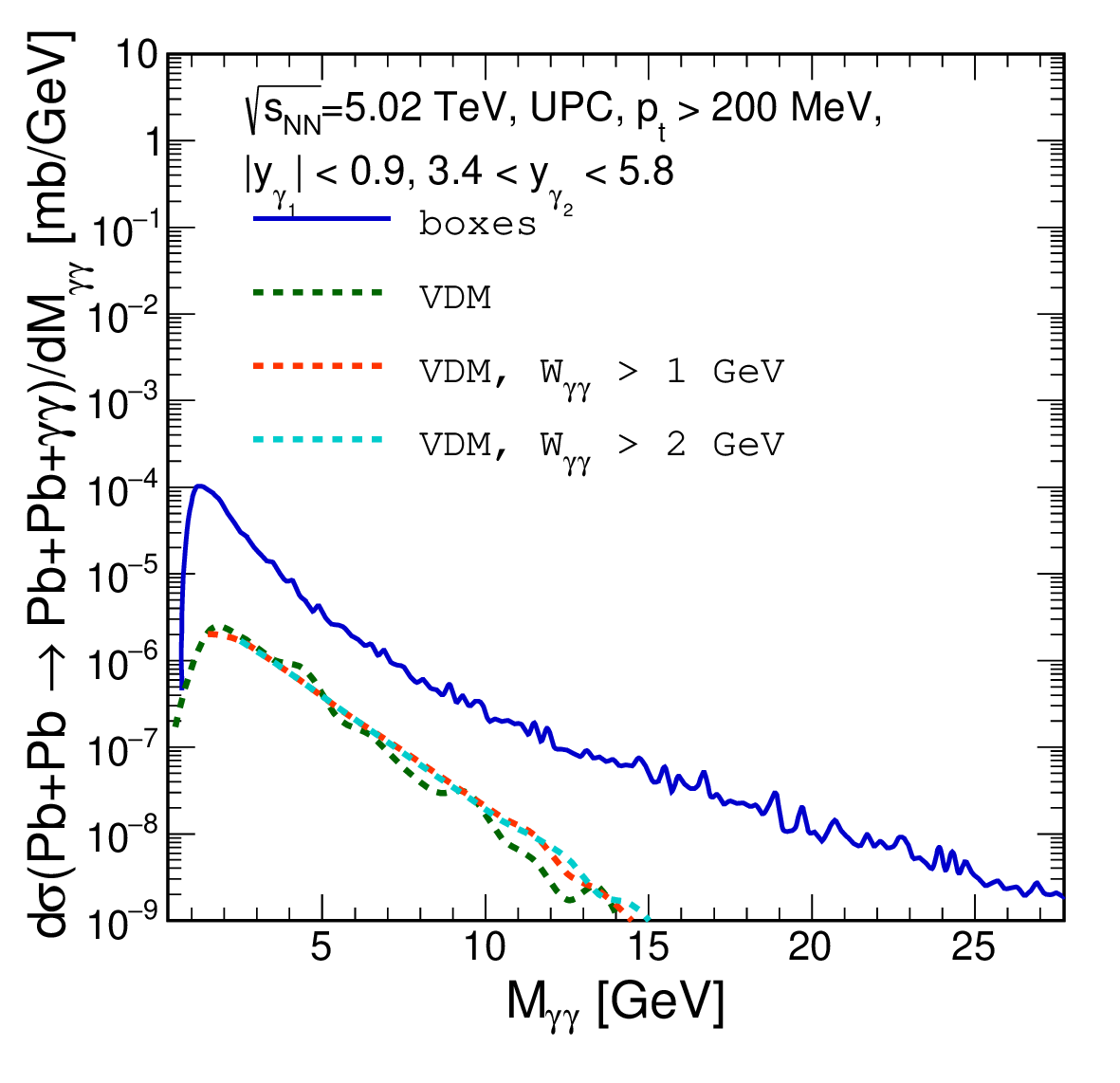}
	(b)\includegraphics[scale=0.3]{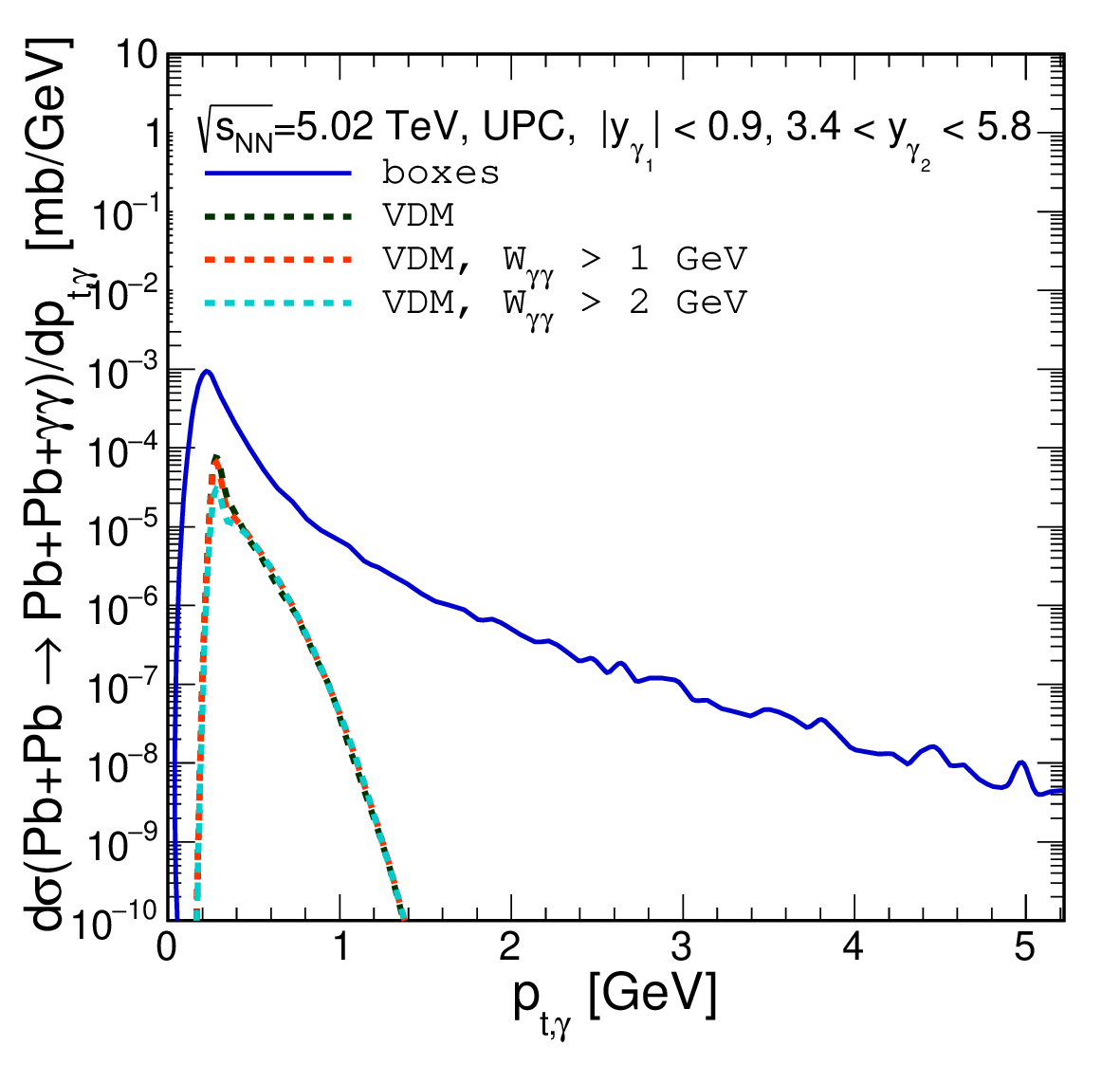}
	\caption{\label{fig:15} Prediction for the FoCal detector in association with mid-rapidity ALICE detector for photons: $p_{t}>200$~MeV, di-photon mass $M_{\gamma\gamma}$ $>$ 400~MeV and photon rapidities $|y_{1}|< 0.9$ and $y_{2} \in (3.4,5.8)$. The blue line corresponds to fermionic loops and the green lines to the VDM-Regge contribution. (a) Di-photon invariant mass, (b) photon transverse momentum distribution.}
\end{figure}

In Fig.\ref{fig:15} we show similar distributions but for $p_t$~$>$~0.2~GeV and combined ALICE and FoCal rapidity region. Here in some regions of the phase space, the VDM-Regge contribution could be seen as $~10\%$ modification of the cross section with respect to the calculations with only boxes.
Here the separated VDM-Regge component is even bigger. We conclude that already at Run 4 one could indirectly observe a signature of other mechanisms than fermionic boxes.

\subsubsection{ALICE 3 kinematics}

Now we wish to show distributions relevant for the ALICE 3 detector.

\begin{figure}[!h]
	\includegraphics[scale=0.32]{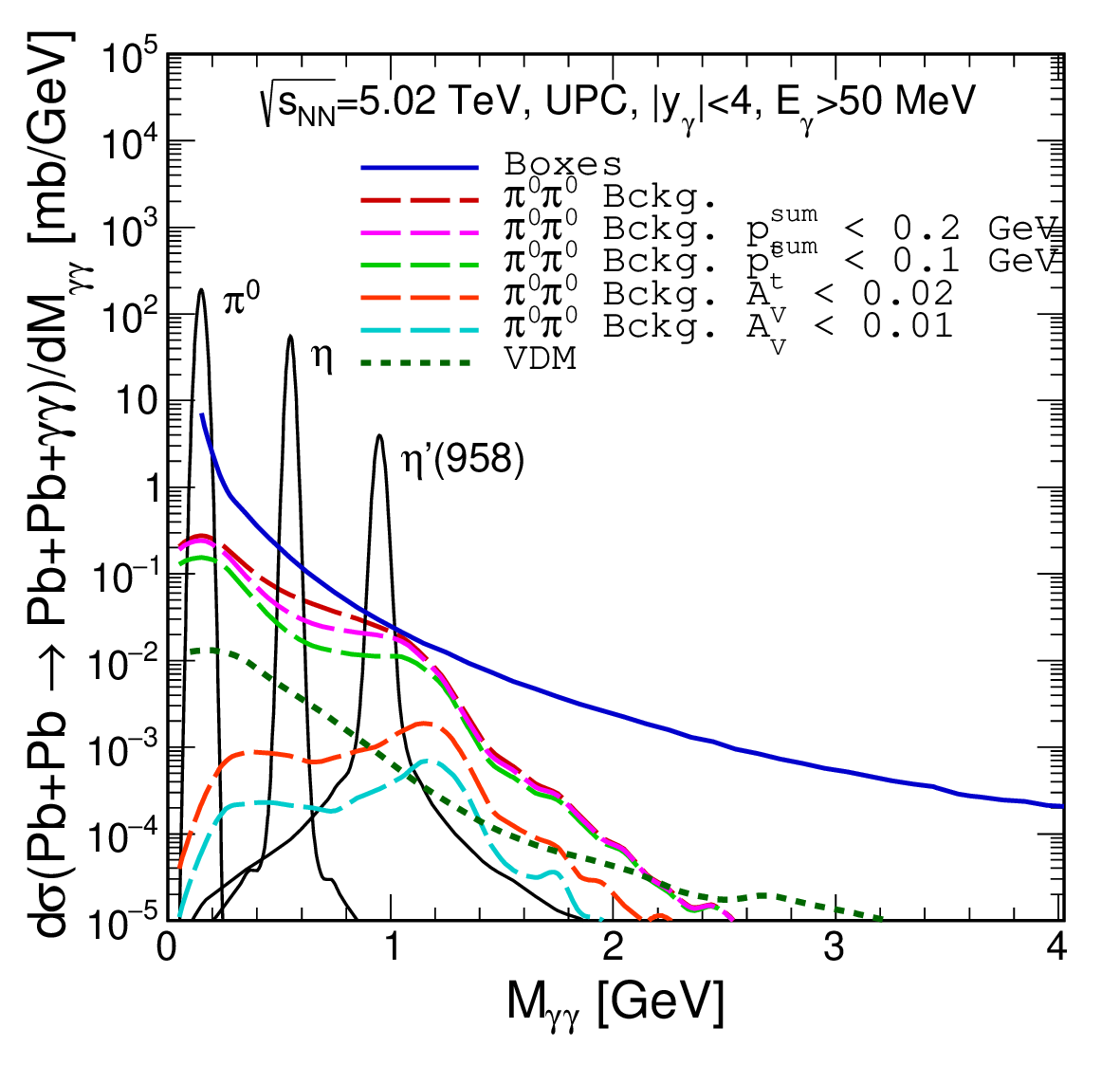}
	\caption{\label{fig:ALICE3_m} Di-photon invariant mass distribution for ALICE 3, i.e. rapidity $y_\gamma \in (-4,4)$ and photon energy E$_{\gamma}>50$~MeV. Here the blue solid line relates to the box contribution, the dotted line to the VDM-Regge component and the dashed lines are for double-$\pi^0$ background contribution. Here we impose several extra conditions on di-photon transverse momenta and vector asymmetry.}
\end{figure}

In Fig.~\ref{fig:ALICE3_m} we show distributions in di-photon
invariant mass for photons $-4< y_1, y_2 <4$ and $E_{\gamma} >50$~MeV (see Ref.~\cite{ALICE3}). We show the light-by-light box
contribution (solid line) as well as the $\pi^0 \pi^0$ background 
contribution (red lines).
At di-photon invariant masses, $0.5$~GeV~$<M_{\gamma\gamma}<1$~GeV, the background contribution is almost
as big as the signal contribution. As discussed in \cite{KSS}
it can be to some extent reduced. Although the background is smaller than fermionic boxes in the full range of di-photon invariant mass, it can be further reduced by imposing the cut on $|\vec{p_{1t}} + \vec{p_{2t}}|<0.1$~GeV and vector asymmetry $A_V<0.02$. Imposing a limit on the background causes that the background in the whole di-photon invariant mass range is much smaller than the signal.


\begin{figure}[!h]
	\includegraphics[scale=0.32]{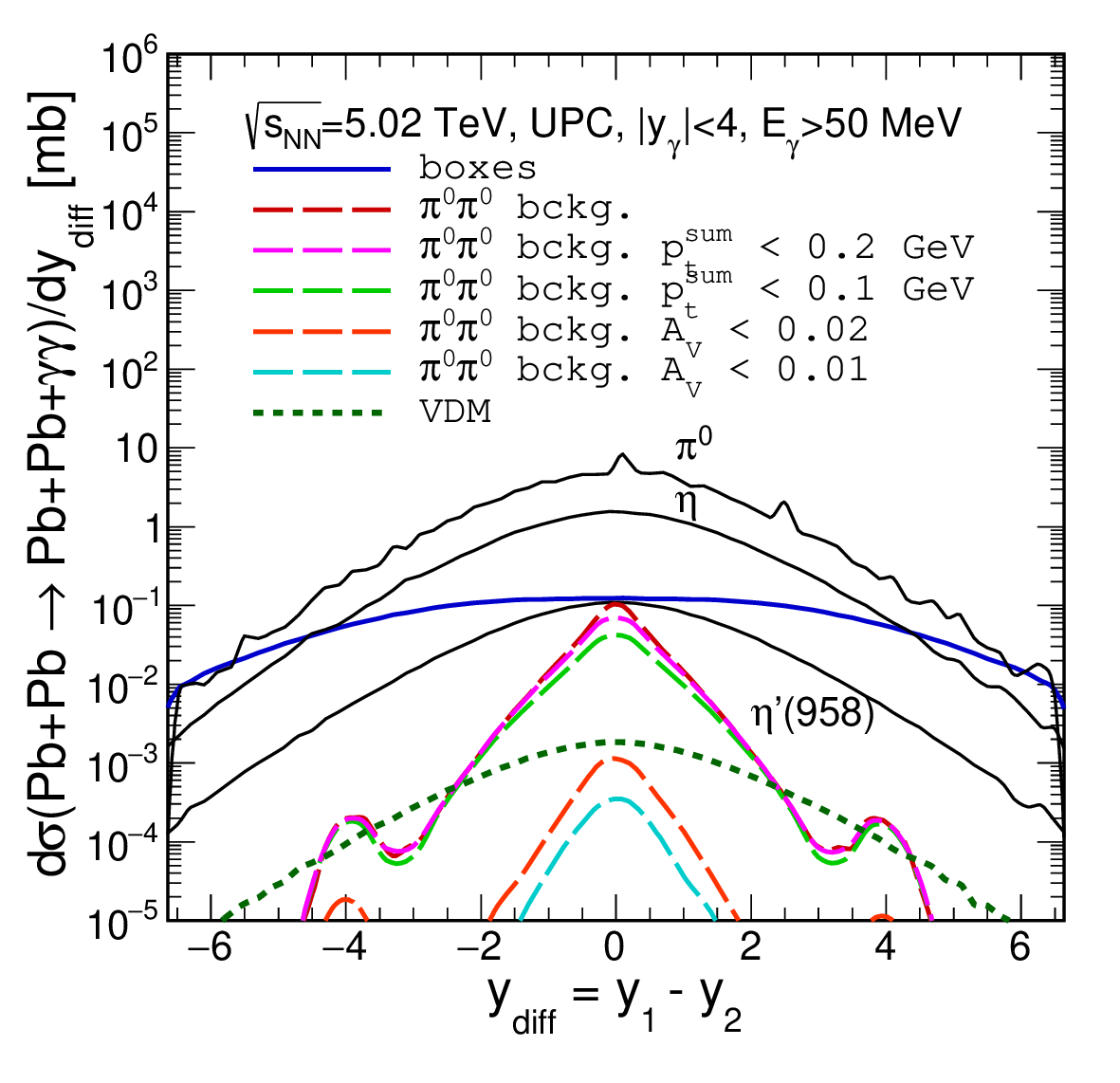}
	\caption{\label{fig:ALICE3_ydiff} Differential cross section as a function of $y_{diff} = y_1 - y_2$ for extended ALICE 3 kinematics: $|y_\gamma|<4$ and $E_\gamma>50$~MeV. Results are presented for boxes, resonances, VDM-Regge and double-$\pi^0$ background.}
\end{figure}

\begin{figure}[!h]
	\includegraphics[scale=0.32]{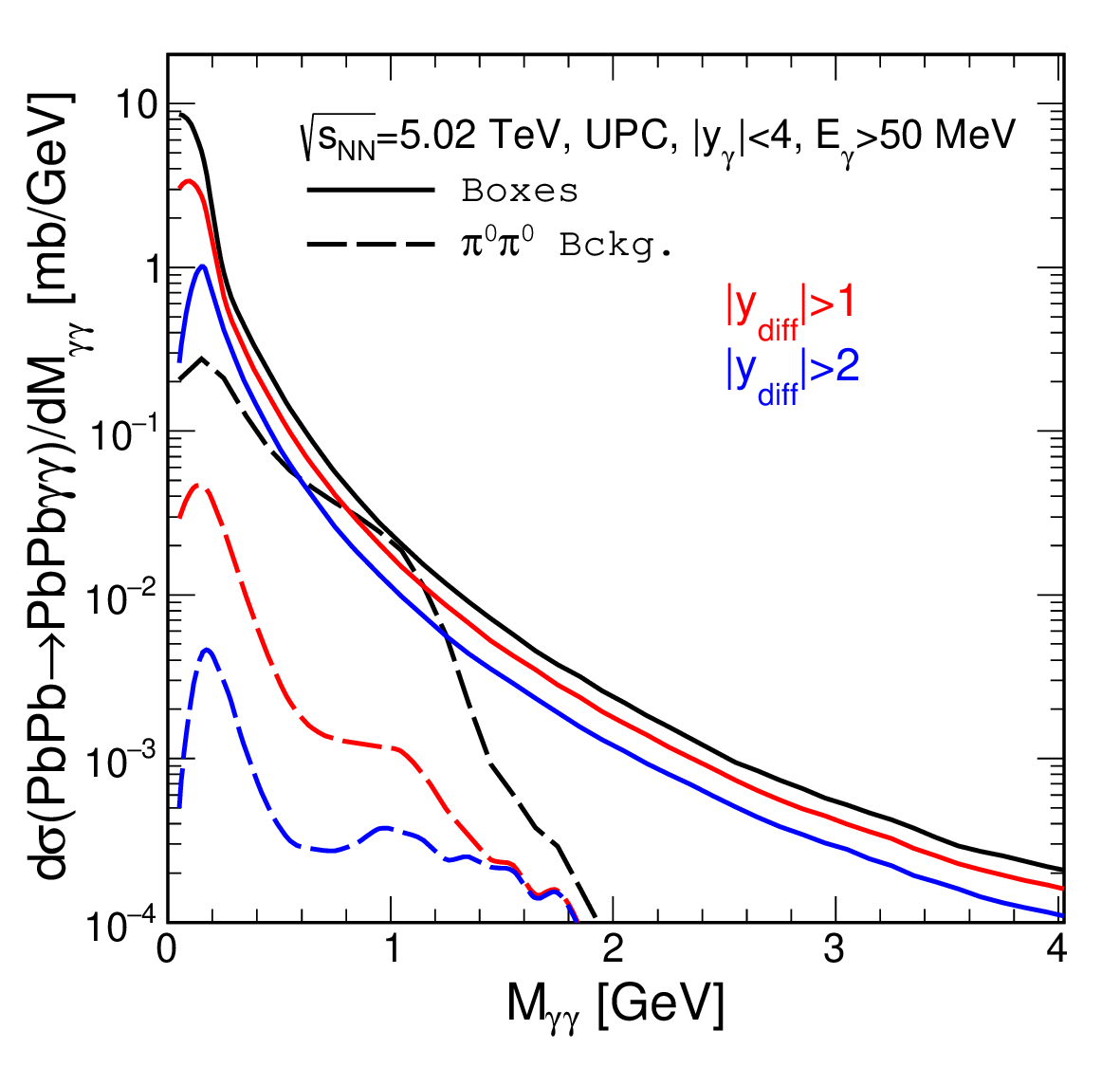}
	\caption{\label{fig:ALICE3_m_ydiff} Influence of extra conditions on $y_{diff} = y_1-y_2$ ($|y_{diff}|>1, 2$) on di-photon invariant mass distribution for ALICE 3. Here the solid lines relate to the box contribution and the dashed lines are for double-$\pi^0$ background contribution.}
\end{figure}

In Fig.~\ref{fig:ALICE3_ydiff} we show analogous distribution in $y_{diff}=y_1-y_2$. Again different contributions are shown separately. The results for the double-$\pi^0$ background contribution are particularly interesting. It has a maximal contribution at $y_{diff}=0$ and drops quickly for larger $|y_{diff}|$. An extra cut on $y_{diff}$ could therefore considerably reduce the unwanted double-$\pi^0$ contribution. In Fig.~\ref{fig:ALICE3_m_ydiff} we show what happens when we impose the cut on $y_{diff}$. The effect of such a cut on box contribution is relatively small but leads to huge reduction of the background. The effect of the cut is much larger for small $M_{\gamma\gamma}$ and therefore should be avoided if one is interested in this region of energies.

\begin{figure}[!h]
	(a)\includegraphics[scale=0.32]{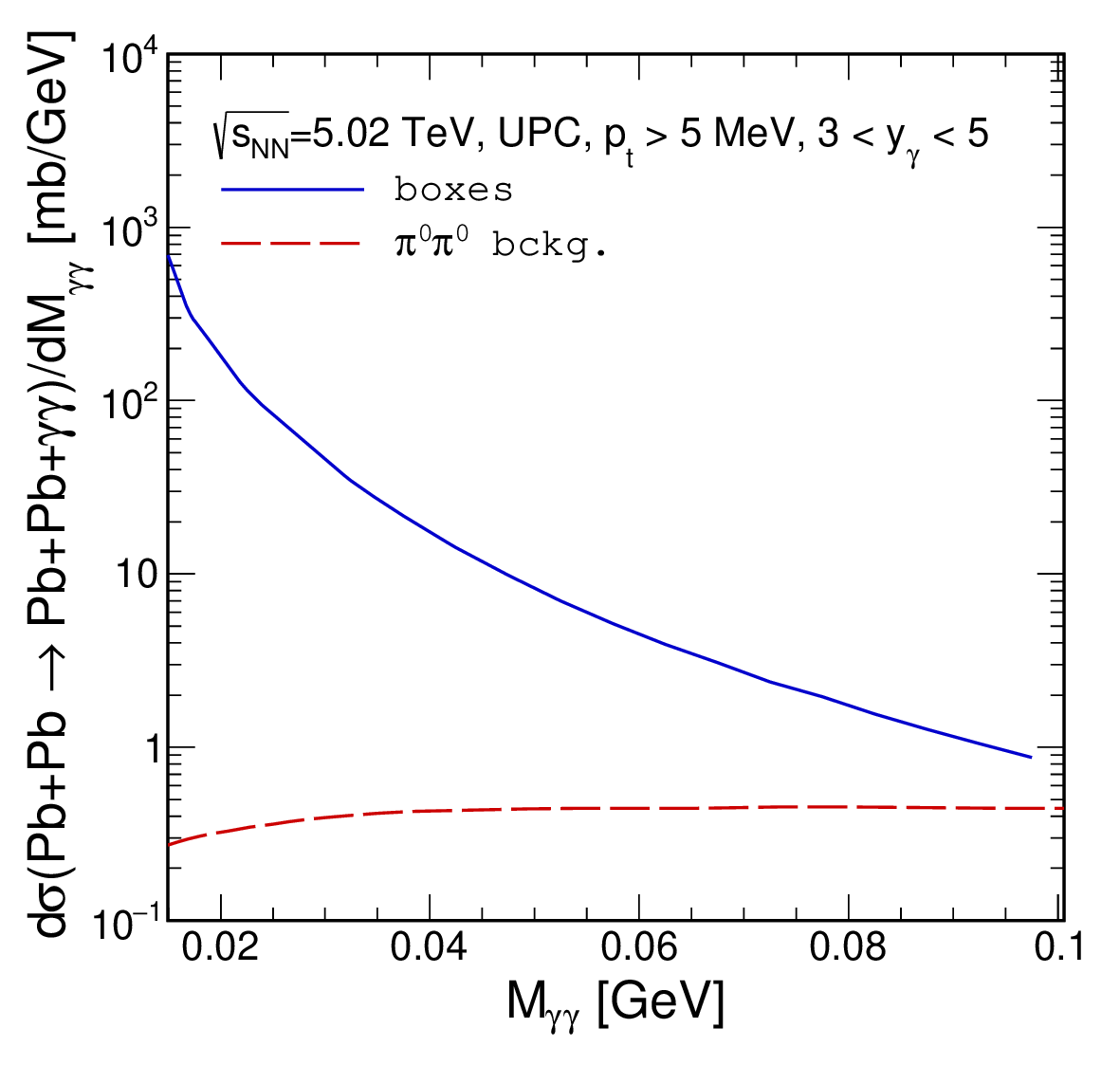}
	(b)\includegraphics[scale=0.32]{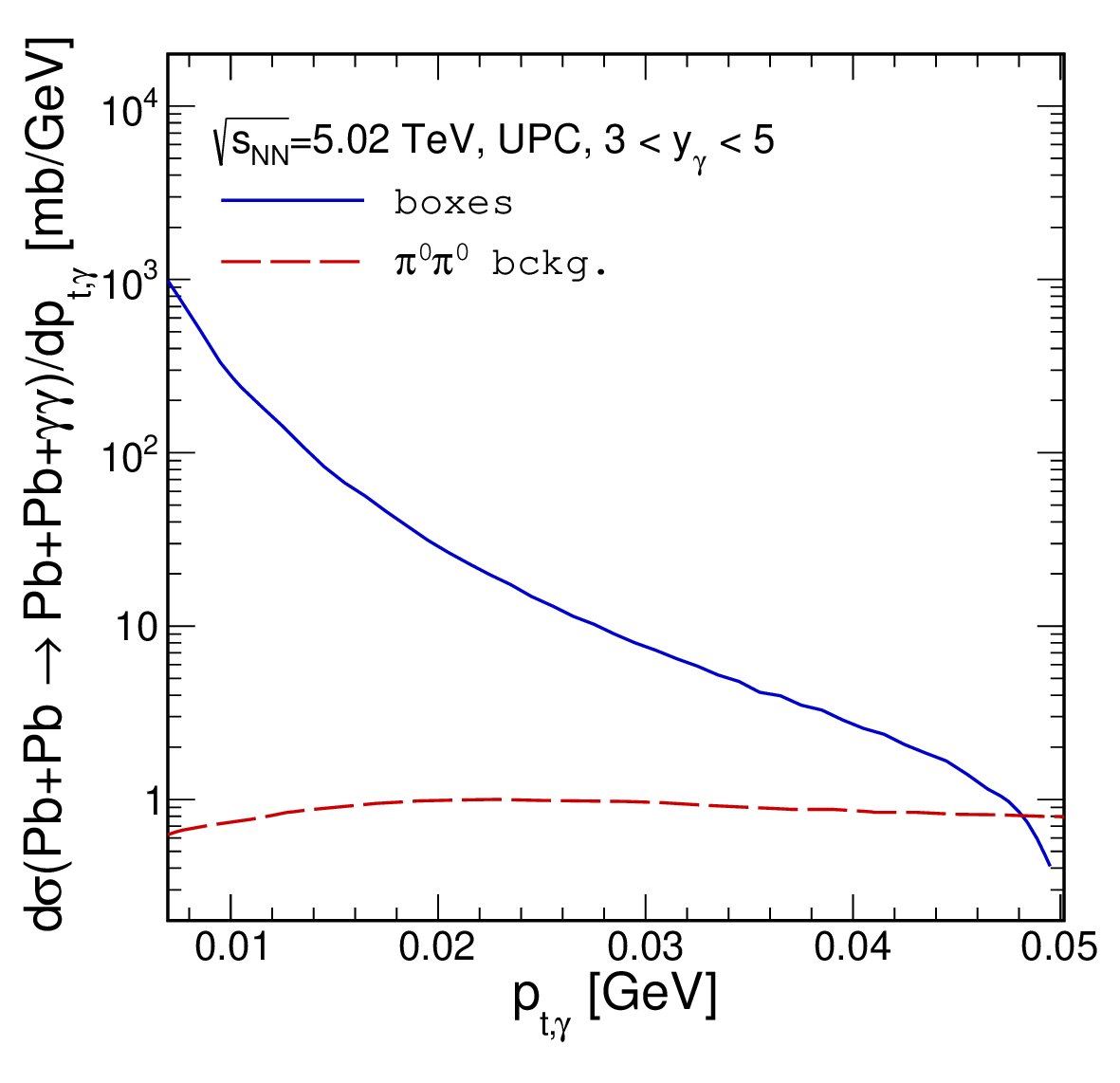}
	\caption{\label{fig:13} Prediction for the ALICE 3 experiment for soft photons: $p_{t} = (5-50)$~MeV and photon rapidities $y_i \in (3,5)$ . The blue line corresponds to fermionic loops, the red line relates to the double-$\pi^0$ background. (a) di-photon invariant mass distribution, (b) photon transverse momentum distribution.}
\end{figure}

In Fig.\ref{fig:13} we show
distribution in $M_{\gamma \gamma}$ (a) and $p_t$ (b)
for a planned special photon detector $3< y_\gamma <5$. Here $p_t >5$~MeV was
imposed as described in Ref.~\cite{ALICE3}. We show that at low
$M_{\gamma\gamma}$ and low $p_t$ the LbL signal by far exceeds the
$\pi^0 \pi^0$ background, even without including any background suppression
condition. Here we have assumed 2 $\pi$ azimuthal coverage of the
special photon detector.

In principle, there is an option to leave the FoCal detector \cite{FoCAL} when running the ALICE 3 detector. This option will be explored elsewhere.

\begin{figure}[!h]
	(a)\includegraphics[scale=0.32]{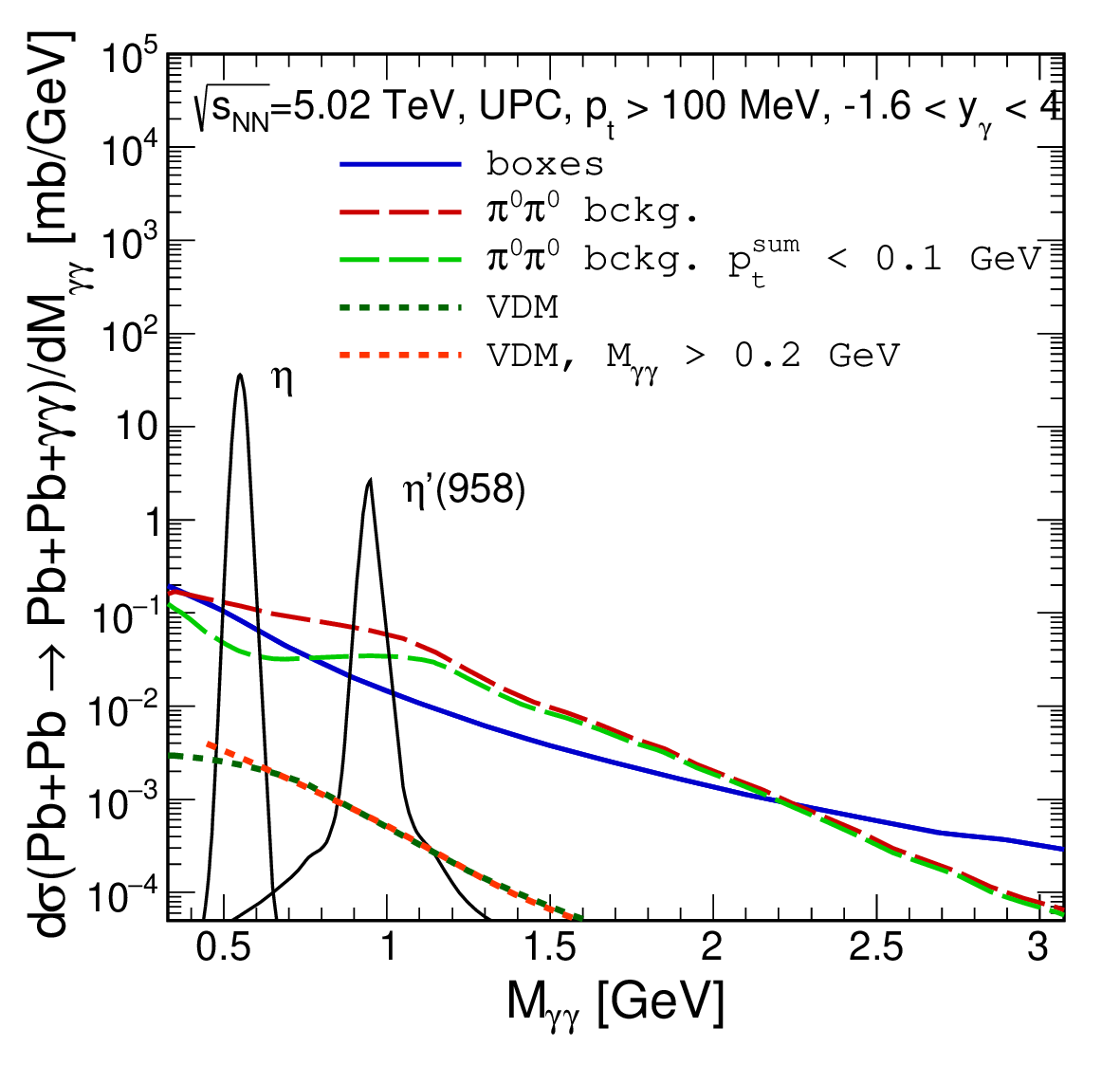}
	(b)\includegraphics[scale=0.32]{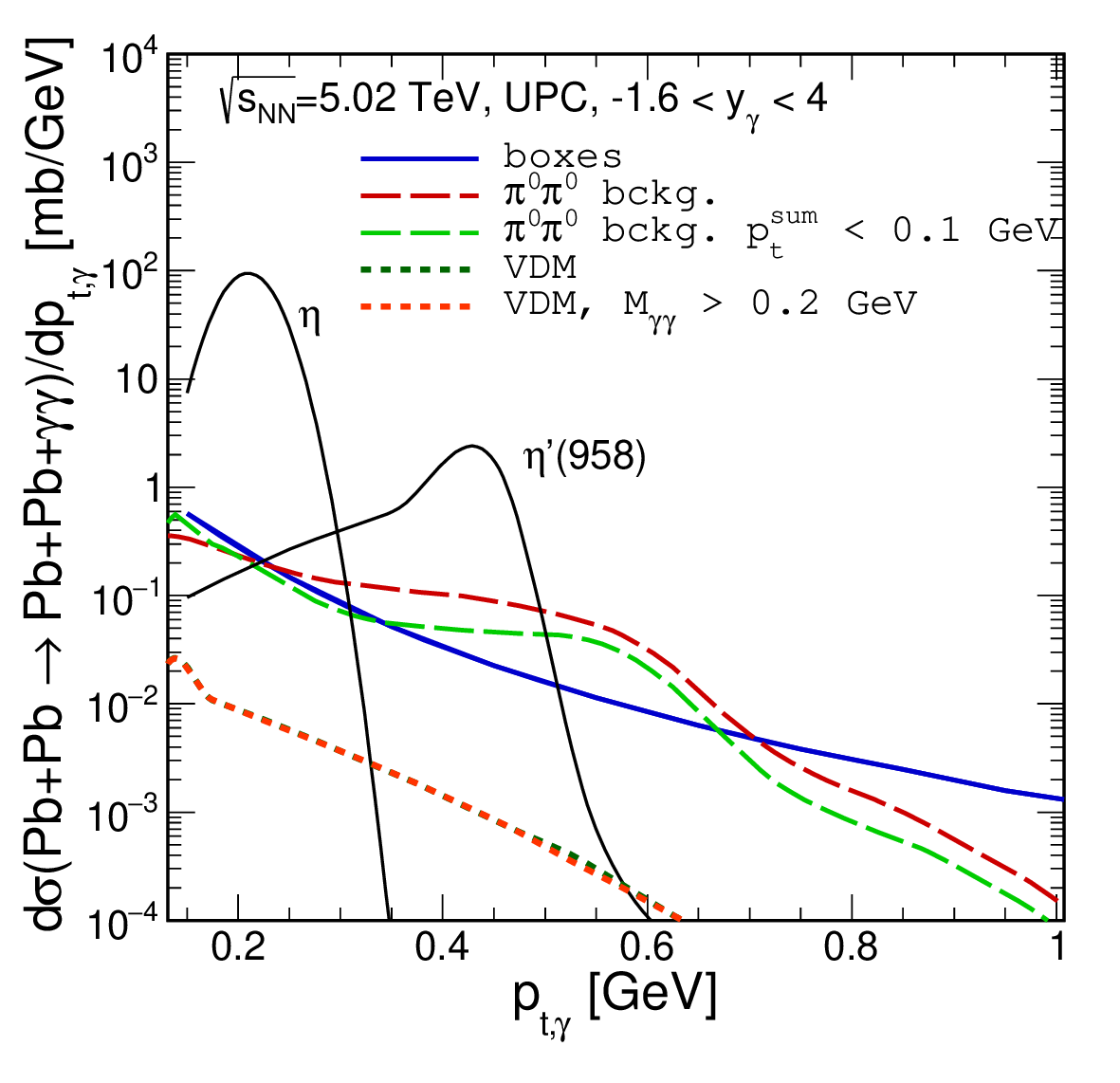}
	\caption{\label{fig:14} 
		Prediction for the ALICE 3 experiment for photons: $p_{t}>100$~MeV and photon rapidities $y_i \in (-1.6,4)$. The blue line corresponds to fermionic loops, the red line relates to the double-$\pi^0$ background. (a) Di-photon invariant mass, (b) photon transverse momentum distribution. }
\end{figure}

At somewhat larger $M_{\gamma \gamma}$ and/or $p_t$ the background
contribution becomes as big as the signal (box) contribution
(see Fig.\ref{fig:14}).
Imposing a cut $|\vec{p}_{1,t}+\vec{p}_{2,t}| <$ 0.1 GeV reduces the background contribution.
It can be further reduced by imposing a cut on so-called vector asymmetry $A_V$ (see \cite{KSS}). However, assuming no background constraints, the cross section for this rapidity range is twice as large as the signal contribution, see Tab.~\ref{tab3}. The comparison of the total cross-sections for the three scenarios shows that the most optimistic scenario is the photon forward detection and $p_{t,\gamma} = (1-50)$~MeV. Here, the double-$\pi^0$ background does not play significant role.

\begin{table}[!h]
	\begin{tabular}{l c r| r| r} 
		\hline
		\multicolumn{3}{c|}{kinematical limitation}  & $\sigma(\pi^0\pi^0$ bckg) $\mu b$ & $\sigma($boxes) $\mu b$\\ 
		\hline
		$p_t>$5 MeV & $E_{\gamma}>50$~MeV & $-4<\eta_\gamma<4$ & 112.824 & 952.590\\
		$p_t>100$ MeV & $E_{\gamma}>50$~MeV & $-1.6<\eta_\gamma<4$ & 159.231 & 82.682\\
		$p_t = (1-50)$ MeV & $E_{\gamma}>50$~MeV & $3<\eta_\gamma<5$    & 105.301 & 3095.795\\
		\hline
	\end{tabular}
	\caption{Fermionic box signal contribution versus double-$\pi^0$ background given in $\mu$b at PbPb collision energy $\sqrt{s_{NN}} = 5.02$~TeV, with three considered scenarios of ALICE 3 detector limitations.}
	\label{tab3}
\end{table}

\section{\label{sec:level5} Conclusions}

We have discussed different mechanisms of $\gamma \gamma \to \gamma \gamma$
scattering such as leptonic/quarkish boxes, double hadronic
fluctuations, neutral $t/u$-channel pion exchanges and two-gluon exchanges. Possible
effects of the subleading mechanisms have been discussed. The latter contributions turned out
difficult to be identified in previous ATLAS and CMS measurements.
We have discussed possible interference effect of box and double-hadronic fluctuations for $\gamma\gamma \to \gamma\gamma$ scattering.

We have calculated several differential distributions for diphotons
in the equivalent photon approximation such as $d \sigma / d p_t$,
$d \sigma / d y$, $d \sigma / d M_{\gamma \gamma}$, $d \sigma / d y_{diff}$,
$d \sigma/dy_1 dy_2$ for PbPb$\to$PbPb$\gamma \gamma$.
Imposing cuts on $y_{diff}$ and transverse momenta of each of the photons
($p_t <$ 0.2 GeV) or $M_{\gamma \gamma} >$ 20 GeV and cut on transverse
momenta of photons we have found regions of the phase space with 
the dominance of the hadronic excitation of both photons, called 
in this paper VDM-Regge for brevity.
Identification of this mechanism at ALICE 3 would be very interesting
and would allow to verify the concept of the VDM-Regge experimentally.

We have also considered the two-$\pi^0$ background, one photon from one
$\pi^0$ and a second photon from the other $\pi^0$,
which gives a relatively large cross section which could be misidentified as the $\gamma\gamma \to \gamma \gamma$ process. 
Imposing lower cuts on $|\vec{p}_{1t} + \vec{p}_{2t}|$ or 
alternatively azimuthal angle between photons eliminates a big part 
of the unwanted two-$pi^0$ background but still some contribution at $M_{\gamma\gamma} \sim 1$~GeV survives.

We have also explored an option to use the planned FoCal detector. When used simultaneously with the ALICE main detector it may allow to study the $\gamma\gamma\to\gamma\gamma$ scattering for $W_{\gamma\gamma}$~$<$~1~GeV, a new unexplored region of the subsystem energies.
Therefore we conclude that the measurement of 
$\gamma + \gamma \to \gamma + \gamma$ in a unique region of relatively 
small energies and small transverse momenta (much lower than those for ATLAS or CMS) will be possible.
 
We have made also predictions for the ALICE 3 ($-4<y_\gamma<4$) and for a planned special soft photon detector ($3<y_\gamma<5$). We have shown that imposing a cut on $y_{diff}=y_1-y_2$ one can eliminate the unwanted double $\pi^0$ background. The soft photon detector can be used to measure the $\gamma\gamma\to\gamma\gamma$ scattering at extremely small energies, $W_{\gamma\gamma}<0.05$~GeV.  
Therefore we conclude that the ALICE 3 infrastructure will be an extremely useful
to study the $\gamma \gamma \to \gamma \gamma$ scattering in a new,
not yet explored, domain of energies and transverse momenta. In this
domain the double-$\pi^0$ background can to large extent be eliminated.

For photon-photon energies (diphoton invariant masses)
1 GeV $ < W < $ 2 GeV the tensor meson exchanges may play an important
role (see \cite{LP}). They would lead to deviations from the box
component. This effect may be difficult to eliminate by imposing
cuts on diphoton invariant mass.

In the present calculations we used EPA in the impact parameter space.
In the future one can try to use also so-called Wigner function approach (never
used for the di-photon production).
This goes, however, beyond the scope of the present exploratory paper.

\section{Acknowledgments}

This work was partially supported by the Polish National Science Center grant UMO-2018/31/B/ST2/03537 and by the Center for Innovation and Transfer of Natural Sciences and Engineering Knowledge in Rzeszów.

We are indebted to Adam Matyja and Jacek Otwinowski for a discussion on ALICE 3 and FoCAL projects. 

\nocite{*}
\bibliography{aipsamp}

\end{document}